\documentclass[aps,pra,twocolumn,reprint,noeprint,superscriptaddress,amsmath,amssymb]{revtex4-1}
\usepackage{graphicx}
\usepackage{bm}
\usepackage{color}
\usepackage{subfigure}
\usepackage{ulem}
\usepackage{amsmath}
\usepackage{amssymb}
\usepackage{mathrsfs}
\usepackage[colorlinks,linkcolor=magenta,anchorcolor=blue,
            citecolor=blue]{hyperref}% add hypertext capabilities
\usepackage[toc,page,title,titletoc,header]{appendix}
\newcommand{\br}[0]{\mathbf{r}}

\def\Rb87{^{87}\rm{Rb}}					% Rb 87
\def\Na23{^{23}\rm{Na}}					%
\begin{document}
\title{Universal driven critical dynamics across a quantum phase transition in ferromagnetic spinor atomic Bose-Einstein condensates}

\author{Ming Xue}
\affiliation{State Key Laboratory of Low Dimensional Quantum Physics, Department of Physics, Tsinghua University, Beijing 100084, China}

\author{Shuai Yin}
\affiliation{Institute for Advanced Study, Tsinghua University,  Beijing 100084, China}

\author{Li You}
 \affiliation{State Key Laboratory of Low Dimensional Quantum Physics, Department of Physics, Tsinghua University, Beijing 100084, China}
 \affiliation{Collaborative Innovation Center of Quantum Matter, Beijing 100084, China}

\date{\today}

\begin{abstract}
We study the equilibrium and dynamical properties of a ferromagnetic spinor atomic Bose-Einstein condensate. In the vicinity of the critical point for a continuous quantum phase transition, universal behaviors are observed
both in the equilibrium state and in the dynamics when the quadratic Zeeman shift is swept linearly.
Three distinct dynamical regions are identified for different sweeping time scales ($\tau$), when compared to the time scale $\tau_{\rm KZ}\sim N^{(1+\nu z)/\nu d}$ decided by external driving in a system with finite size $N$ ($\nu,z$ are critical exponents and $d$ the dimensionality).
 They are manifested by the excitation probability $\mathcal{P}$ and the excess heat density $\mathcal{Q}$. The adiabatic region of $\,\mathcal{P}\sim\mathcal{Q}\sim\tau^{-2}\,$ follows from the adiabatic perturbation theory when $\tau >\tau_{\rm KZ}$, while the non-adiabatic universal region of $\,\mathcal{P}\sim\mathcal{Q}\sim\tau^{-1}\,$ in the thermodynamic limit
is described by the Kibble-Zurek mechanism when $\tau_{\rm KZ}>\tau >$ the time scale given by initial gap. The Kibble-Zurek scaling hypothesis is augmented with finite-size scaling in the latter region
and several experimentally falsifiable features for the finite system we consider are predicted.
The region of the fastest sweeping is found to be non-universal and far-from-equilibrium with $\mathcal{P}$ and $\mathcal{Q}$ essentially being constants independent of $\tau$.
\end{abstract}
\maketitle
%\tableofcontents
\section{Introduction}
Discovering and understanding nonequilibrium scaling behaviors near the quantum critical point (QCP) is one of the most interesting arenas in condensed matter physics and statistical physics. Continuous quantum phase transitions (QPTs) occur when the control parameter in a Hamiltonian is tuned across QCPs at zero temperature \cite{sachdev2011quantum}.
In a continuous phase transition, the order parameter vanishes smoothly as the critical point is approached.
The existence of a QCP is usually accompanied by nonanalyticity in the ground state energy, and it usually connects two quantum phases with different symmetries. Strong quantum fluctuations near a QCP always lead to breaking of symmetry and subsequent building up a macroscopic order.  The emergence of an order parameter and the nonanalyticity in the ground state energy are related by the Hellmann-Feynman theorem.

Universality, which originates from the scale invariance near a critical point, is a remarkable feature in continuous phase transitions \cite{cardy1996scaling,Stanley1999}. As is known from equilibrium critical phenomena in classical systems, universal behaviors emerge in the vicinity of a critical point where a large number of degrees of freedom are strongly correlated. Associated with the critical point a set of critical exponents can be used to describe the scaling behaviors for relevant quantities near the transition. Moreover, the classical notion of universality in thermal phase transition has been extended successfully to describe the quantum critical phenomena due to quantum fluctuations at zero temperature \cite{sachdev2011quantum}.

Cold atom experiments facilitate the study of quantum phases and their associated QPTs in a closed quantum many-body system \cite{Bloch2008,Polkovnikov2011,
stamperkurn2013spinor,Langen2015}.
A wide variety of dynamical properties can be monitored
because the relevant energy scales in cold atom systems are much smaller than in conventional condensed matter systems, therefore the relaxation time or the response time is longer and easier to follow experimentally.
The equilibrium relaxation time $t_\text{eq}$ of a quantum system, which is typically measured by the  inverse of the excitation gap ($\Delta$), diverges in the thermodynamic limit (TDL) because of the gap closing at the QCP. Consequently any driving of the control parameter at a finite rate would cause nonequilibrium effects. An effective approach for the description of such nonequilbrium effects is the celebrated Kibble-Zurek (KZ) mechanism  \cite{kibble1976topology,zurek1985cosmological,zurek1996cosmological}, which was first proposed in cosmology physics by Kibble and then extended by Zurek to condensed matter physics.

\par The KZ mechanism has been extensively studied both in classical and quantum systems, and in theories \cite{
Damski2005,Zurek2005,Damski2006,Damski2007,Lamacraft2007,Saito2007,Uwe2007,Cucchietti2007,
DelCampo2010,Sabbatini2011,Saito2013a,Huang2014,Lee2015,
Jaschke2016} as well as in experiments \cite{chuang1991cosmology,bauerle1996laboratory,ruutu1996vortex,
Chen2011,Baumann2011,Lamporesi2013,Corman2014,
Navon2015,Clark2016,Anquez2016,Aidelsburger2017a}. A signature scaling relation between the number of defects or excitations and the driving rate is predicted when the system is driven across a continuous phase transition. The key enabling element lies at the possibility of combining the equilibrium critical exponents and the driving rate to characterize the nonequilibrium effects from the finite driving rate. The main idea involves seperating the whole dynamics in such a driven process into an adiabatic plus an impulse region.  When the driven parameter is far from the critical point, the dynamics is approximately adiabatic due to large equilibrium relaxation time; When the critical point is approached, due to the so-called critical slowing down, the system dynamics can be regarded as frozen and describable by the impulse approximation, and nonadiabatic effects appear. The instant separating the two regions is obtained by equating the time remained to arrive at the QCP, denoted as $t_{\rm KZ}$, to the equilibrium relaxation time $t_{\rm eq}$, {\it i.e.,} $t_{\rm KZ}\simeq t_\text{eq}\simeq {1}/{\Delta}$. The different dynamic regions then originate from the competitions between the two time (length) scales \cite{Huang2014}: the time (length) scale given by external driving and intrinsic relaxation time $t_\text{eq}$ (correlation length $\xi$).

Spinor atomic Bose-Einstein condensate (BEC) exhibits rich magnetic phases in the presence of external magnetic field, which makes it a suitable platform to study the dynamics of QPTs. In this work, we focus on a  spin-1 BEC with ferromagnetic interactions such as for $\Rb87$ atoms  \cite{Stenger1998,Barrett2001,Chang2004,Sadler2006,Luo2017}.
Invariably, current atomic BEC systems are trapped in a finite volume by magnetic or optical means with a finite number of atoms, although the total atom number can be changed to some degree from experiment to experiment.
 In the pioneering experimental work of Ref.\,\cite{Anquez2016}, aimed at checking the predictions of KZ mechanism, the scaling behavior for the impulse stage duration
was confirmed. But the deviation of the scaling exponent from the mean-field theory critical exponent is evident especially at the long ramp time limit. It is presumably due to the neglect of the finite size effect, which enters by opening a gap at the QCP and smoothes out the relevant phase transition observables. It cannot be ignored especially when the finite gap opening at the QCP is comparable with the energy scale associated with the dynamics one is investigating. Besides, a finite gap enables near-adiabatic preparation of metrologically meaningful quantum states \cite{Luo2017}.

The equilibrium and dynamical properties are studied in this work when the quadratic Zeeman shift is tuned through a continuous QCP as in recent experiments \cite{Anquez2016,Luo2017}.
We combine the KZ mechanism with finite-size scaling theory to obtain universal dynamical scaling functions for relevant phase transition observables and successfully verify their scaling collapse in finite systems by using the mean-field critical exponents.
We cover the whole range of the driving rate and find that the dynamics in a finite system can be described by adiabatic perturbation theory  \cite{Polkovnikov2008a,DeGrandi2010} in the very slow driving limit, and becomes far-from-equilibrium and non-universal in the fast driving limit.

\par This paper is organized as follows. We first discuss the QPT for our model in Sect.\,\ref{subsec:modelHam} and extract the critical exponents from mean-field results in Sect.\,\ref{subsec:exponents}. In Sect.\,\ref{subsec:FSSeq}, we study the finite-size scaling for equilibrium observables. Section\,\ref{sec:dynamic} is devoted to a study of the dynamical properties for a linear driven protocol, where three distinct dynamical regions are analyzed. The consequent predictions can be tested in existing experimental setups. Finally in Sect.\,\ref{sec:conclusion}, we conclude with discussions.\\

\begin{figure}%[!htbp]
  \centering
 \includegraphics[width=0.8\columnwidth]{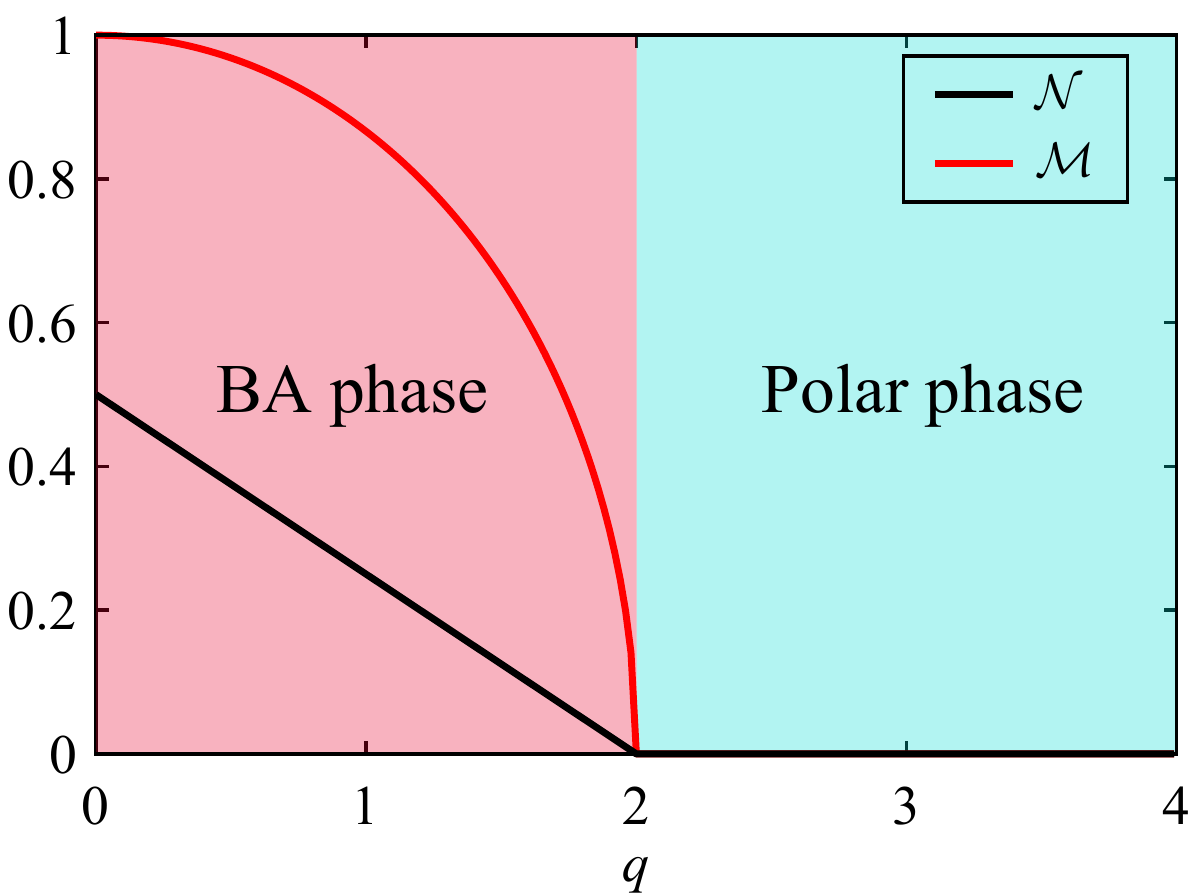}
  \caption{{\bf Mean-field phase diagram.} The mean-field phase diagram for our model at $q > 0$ in the subspace of zero longitudinal magnetization $F_z = 0$. The broken-axisymmetry phase (BA phase) and the polar phase are separated at the quantum critical point(QCP) $q_c = 2$. The mean-field values for ground state observables: fractional population $\mathcal{N}$ (black solid line) and transverse magnetization $\mathcal{M}$ (red solid line). The mean-field critical exponents can be obtained from the scaling behaviors near the QCP for $\mathcal{N}$ and $\mathcal{M}$ (see the main text).}\label{fig:phasediag}
\end{figure}
\section{Model Hamiltonian and the Critical exponents}\label{sec:modelHam}
\subsection{Spin-1 BEC Hamiltonian and its QPT}\label{subsec:modelHam}
{\it Model.}---For a spin-1 BEC of $\Rb87$ or $\Na23$ atoms, the spin-dependent interaction strength is usually much weaker than the density-density interactions, it is therefore reasonable to make the single-mode approximation (SMA) by assuming that all spin states share the same spatial wavefunction $\phi(\br)$, which is unit normalized according to $\int|\phi(\br)|^2d\br =1$ \cite{Law1998}. SMA decouples the spatial mode and the spin. The equations of motion at low energies are simplified to those concerning the internal spin degrees of freedom. The Hamiltonian under SMA becomes \cite{Law1998,Pu1999}
\begin{widetext}
\begin{eqnarray}
  \hat H &= &\frac{c_2}{2N}\left[\left(2\hat N_0 -1\right)\left(\hat N_1+\hat N_{-1}\right)+2\left(\hat a_1^\dag\hat a_{-1}^\dag\hat a_0\hat a_0+\text{h.c.}\right)\right]
   -p \left(\hat N_1 - \hat N_{-1}\right)
  + \,q\,\left(\hat N_1 + \hat N_{-1}\right)\,,\label{eq:Hamil0}
\end{eqnarray}
\end{widetext}
where $\hat a_{m_f} (m_f=0,\pm1)$ is the annihilation operator of the ground state manifold $|f=1, m_f\rangle$, with number operator $\hat N_{m_f} = \hat a_{m_f}^\dag\hat a_{m_f}$, and the total particle number operator $\hat N = \hat N_{1}+\hat N_0+\hat N_{-1}$ is conserved. $p$ and $q$ are linear and quadratic Zeeman shifts which could be tuned independently in experiments.
The spinor dynamic rate $c_2$, which sets the spin-dependent interaction energy scale, is defined as $c_2 = N\int |\phi(\br)|^4 d\br\times\frac{4\pi (a_2-a_0)}{3m_\text{a}}\,,$ with $m_\text{a}$ being the atomic mass, $a_F$ the $s$-wave scattering length in the total spin angular momentum channel of $F=f_1+f_2$ for the two atoms. Atomic interactions naturally give $c_2<0$ for $\Rb87$ atoms and $c_2>0$ for $\Na23$ atoms which corresponds to ferromagnetic and anti-ferromagnetic spin-dependent interactions, respectively.
\par The collective spin operators for this spin-1 boson system are defined by
$\hat F_+ = \sqrt{2}\,(\hat a_1^\dag \hat a_0+\hat a_0^\dag\hat a_{-1}),\, \hat F_-= \hat F_+^\dag,\,
  \hat F_z =\hat a_1^\dag a_1-\hat a^\dag_{-1}\hat a_{-1},
$
 where $\,\hat F_\pm \equiv \hat F_x \pm i\hat F_y\,$ are the raising and lowering operators, and $[\hat F_z, \hat H]=0$, making the longitudinal magnetization $ F_z$ a good quantum number. Hereafter we constrain to the $F_z = 0$ subspace, which means the linear Zeeman shift can be set to $p=0$, effectively.

{\it Phase diagram.}---In the following discussions, we shall focus on the QPT physics in the ferromagnetic condensate with $c_2<0$ and nonnegative (effective) quadratic Zeeman energy $q \geq 0$. As we can see from Eq.\,(\ref{eq:Hamil0}), in the limit of $q/|c_2|\rightarrow +\infty$, all atoms stay in the single-particle state $|1,0\rangle$, but in the limit of $q/|c_2|\rightarrow 0$, the ferromagnetic interaction term dominates. There must exist a critical point when these two terms are comparable.
 The competition between the ferromagnetic interaction and the quadratic Zeeman energy manifests the system by two phases with different symmetries revealed by their collective spin magnetization. They are the polar phase for $q/|c_2| > 2$ and the broken-axisymmerty (BA) phase for $0\leq q/|c_2|\leq 2$ (see Fig.\,\ref{fig:phasediag} for the phase diagram).

In order to clarify the QCP explicitly, we assume a homogeneous density profile $\phi(\br)=\frac{1}{\sqrt{V}}$ for the condensate, which is a good approximation if the atoms are loaded into a flat trap \cite{Gaunt2013,Chomaz2015,Beugnon2016,Mukherjee2017,Hueck2018}. Therefore $c_2 \propto N\int |\phi(\br)|^4 d\br\propto\frac{N}{V}$. Strictly speaking, phase transitions occur only in the limit of thermodynamics
 $ \lim\limits_{N,V\rightarrow\infty}\frac{N}{V} = \text{const.}\,,\,$
 so $c_2$ is intensive and fixed when we take the TDL. From now on we take $|c_2|=1$ as energy unit in the following discussions. If the system is inhomogeneous in space, such as in a 3D harmonic trap \cite{Anquez2016,Luo2017}, under the Thomas-Fermi approximation, one must take $c_2(N)\propto N^{2/5}$ into consideration to keep the interaction energy per atom fixed when the TDL is taken \cite{Anquez2016}.

\par For a continuous transition associated with spontaneously broken symmetry, order parameters can be defined to identify the QPT. The following two order parameters \cite{Damski2007,Lamacraft2007,Anquez2016}
$$
 \mathcal{N} =  \frac{\langle \hat N_1 +\hat N_{-1}\rangle}{N}, \quad
\mathcal{M} =\frac{\sqrt{\langle\hat F_x^2\rangle+\langle\hat F_y^2\rangle}}{N}
\,,$$
are adopted, wherein $\mathcal{N}$ denotes the fractional atomic population in magnetic states $|1,1\rangle$ and $|1,-1\rangle\,$, and $\mathcal{M}$ is the magnitude of the transverse magnetization for the collective spin. $E_n(q)$ denotes the  $n$-th ($n\in\mathbb{N}$) eigenvalue of $\hat H(q)$, and $e_n(q)\equiv E_n(q)/N$ the energy per particle. By using the Hellman-Feynman theorem, the fractional population satisfies $\mathcal{N}(q) \equiv \frac{1}{N}\left\langle\frac{\partial \hat H(q)}{\partial q}\right\rangle= \frac{\partial  e_0(q)}{\partial q}$ with $e_0$ the ground state energy per particle. From Fig.\,\ref{fig:phasediag}, it is clear that the QCP at $q=2\,$ is a second order transition since the derivative of $e_0$ with respect to $q\,$, namely  $\mathcal{N}(q)\,$, is continuous but the higher order derivatives are discontinuous.

\begin{figure}%[!bht]
  \centering
  \includegraphics[width=0.8\columnwidth]{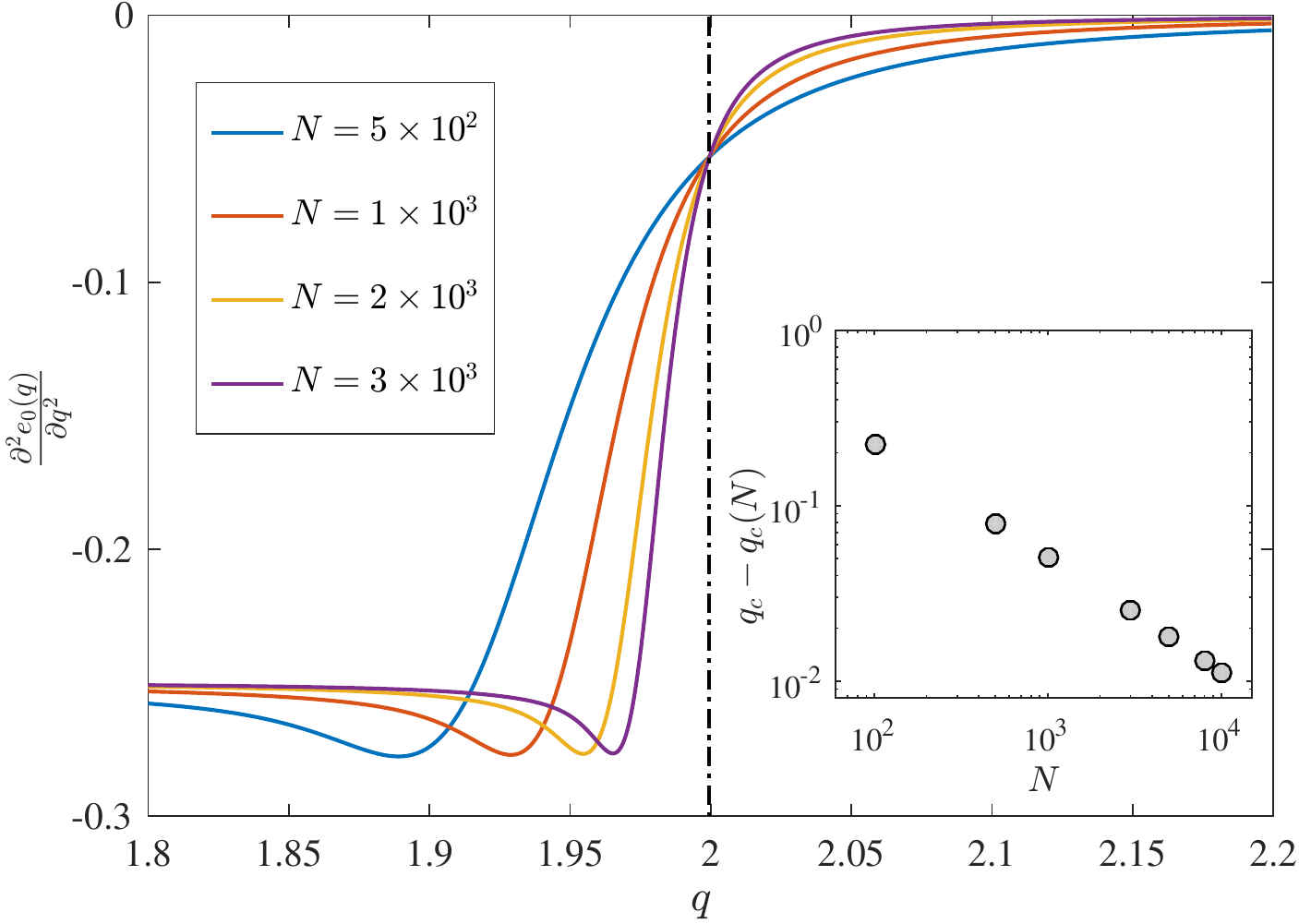}\\
  \caption{{\bf The precursor to QPT in a finite system.} The $\frac{\partial^2 e_0}{\partial q^2}$ approaches a discontinuous step with increasing $N$, which implies a second order (continuous) QPT according to Ehrenfest's classification.  Inset: the pseudo-critical point $q_c(N)$ (location of the minimal $e_1-e_0$) for different finite size $N$. In the log-log plot, the difference $q_c-q_c(N)$ is seeing to vanish as $N\rightarrow\infty$ according to a power law, wherein $q_c=2$ is the mean-field critical point. This indicates the mean-field critical point is exact. }\label{fig:quantumQCP}
\end{figure}
\par Besides the mean-field results,  in Fig.\,\ref{fig:quantumQCP}, we also show numerical results of $\frac{\partial^2 e_0}{\partial q^2}$ obtained from exact diagonalization of the Hamiltonian of Eq.\,(\ref{eq:Hamil0}) for different total atom number $N$.
The increasingly sharper jump from zero to a negative value for $\frac{\partial^2 e_0}{\partial q^2}$ with increasing $N$ serves as a precursor to QPT in a finite system.  The inset of Fig.\,\ref{fig:quantumQCP} shows the locations of the minimal $e_1-e_0$ for different $N$, {\it i.e.,} the pseudo-critical points [$q_c(N)$] for a finite system. It is clear that the $q_c(N)$ converges to $q_c=2$ in the TDL, consistent with the mean-field critical point.

\begin{figure*}[!hbt]
  \centering
  \includegraphics[width = 1.75\columnwidth]{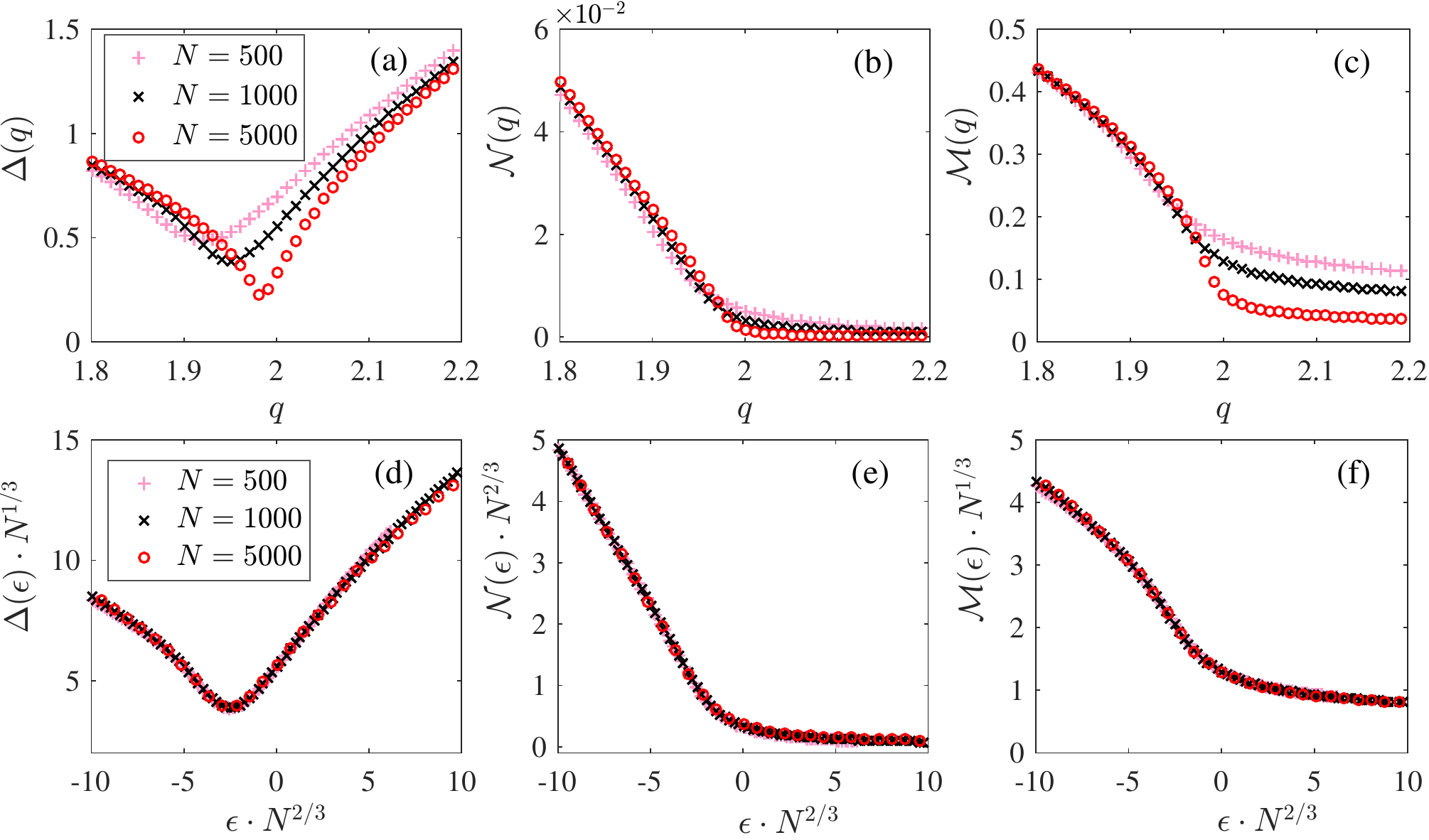}\\
  \caption{{\bf Finite-size scaling at equilibrium.}
  (a)-(c) In the vicinity of the QCP, exact diagonalization of the Hamiltonian of Eq.\,(\ref{eq:Hamil0}) gives the gap $\Delta(q)\,$, fractional population $\mathcal{N}(q)$ and the transverse magnetization $\mathcal{M}(q)$ for the ground state.           (d)-(f) show the corresponding data rescaled according to Eqs.\,(\ref{eq:Gapfss})-(\ref{eq:OPfss}) by using the critical exponents in Table \ref{tab:exponents}.  Finite-size scaling is clearly verified. Different system sizes for $N = 500, 1000 \text{ and } 5000$ are used in the calculations.}\label{fig:rescale_eq}
\end{figure*}
 \subsection{Static critical properties}\label{subsec:exponents}
  The Bogoliubov analysis in Ref.\,\cite{Murata2007} for our model system shows there exist three excitation modes at long wavelength limit in the BA phase. One is gapful and the other two are gapless Goldstone modes associated with U(1)  and SO(2) symmetries being broken. The gapful mode denoted as $E_\alpha$ in Ref.\,\cite{Murata2007} is directly relevant for our following discussions,
\begin{eqnarray*}
  E_\alpha^2 &=& \Delta^2 + 4|c_2|\epsilon_{\mathbf k} + O(\epsilon_\mathbf{k}^2)\;,\\
  \Delta^2 &=& \left(q_c-q\right) \left(q_c+q\right)\;,
\end{eqnarray*}
where $\epsilon_\mathbf{k} = \frac{\hbar^2{\mathbf k}^2}{2m}$ and $\Delta$ are free particle dispersion and excitation gap, respectively.
\par Therefore, the excitation is gapless with a spectrum $E_\alpha\sim\epsilon^{1/2}_\mathbf{k}\sim k^z$ at the QCP $q=q_c$, so we must have the dynamical critical exponent $z=1$. Furthermore, the behavior of the gap approaching the QCP from the BA phase $\Delta({q\rightarrow q_c^{-}})\sim |q-q_c|^{\nu z}$ yields \,$\nu z = 1/2$, thus the correlation length critical exponent $\nu=1/2$.
\par The mean-field results for the order parameters $\mathcal{N}$ and $\mathcal{M}$ near the QCP in the BA phase are respectively given by\,\cite{Murata2007,Hoang2016a},
 \begin{eqnarray*}
  \mathcal{N}^\text{(BA)} &\propto&{q_c-q}\,,\qquad
  \mathcal{M}^\text{(BA)} \propto \sqrt{q_c-q}\;,
\end{eqnarray*}
 as shown in Fig.\,\ref{fig:phasediag}, and both are zero in the polar phase. We thus obtain the exponents of order parameters $\beta_\mathcal{N} = 1$ and $\beta_{\mathcal{M}} = 1/2\,$ from the behavior $\mathcal{O}\sim |q-q_c|^{\beta_\mathcal{O}}$ (where $\mathcal{O} = \mathcal{N}, \mathcal{M}$) in the vicinity of the QCP.

\par The Hamiltonian in Eq.\,(\ref{eq:Hamil0}) actually describes $N$ spin--1 bosons interacting equally with all other spins. For such a system mean-field theory gives exact results about the QPT.
Because of the infinitely long-range nature of interaction, the concepts of ``dimensionality'' or ``length''  are not well-defined \cite{Botet1982,Botet1983}. The correlation length for a general short-range model must be substituted by an effective quantity $N_\xi$.
 By following the arguments of Botet and Jullien \cite{Botet1982,Botet1983}, we can define a length scale $\xi$ which simply links the upper critical dimensionality $d_c$ of the corresponding finite-range model according to $N_\xi\sim \xi^{d_c}$. The finite-range spin model has an upper critical dimension $d_c = 4\,$ for a classical phase transition, and since a QPT in $d$-dimension has the same critical behaviors as the classical transition in $(d+z)$-dimension, the upper critical dimensionality is $d = 4-z=3$ for the QPT we discuss. This dimensionality is consistent with what we have in the approximated Hamiltonian\,(\ref{eq:Hamil0}) under SMA. If the coherence number $N_\xi$ is used as an effective correlation length, we find critical exponents $\nu^\ast z^\ast=1/2$ but with $\nu^* = \nu d = 3/2, \, z^*= z/d=1/3$, which implies the information concerning dimensionality is encapsulated into the critical exponents.
 We list the critical exponents in Table \ref{tab:exponents} for later use.
\begin{table}[!htbp]
\tabcolsep 2pt
\caption{The critical exponents and dimensionality.}
\vspace*{-12pt}
\begin{center}
\def\temptablewidth{0.3\textwidth}
{\rule{\temptablewidth}{1pt}}
\begin{tabular*}{\temptablewidth}{@{\extracolsep{\fill}}cccccc}
   $\;\nu$ & $\beta_\mathcal{N}$ &$\beta_{\mathcal{M}}$ & $z$ & $d$ \, \\   \hline
   $\;1/2$  &   1  & $1/2$ & 1 & 3 \,  \label{tab:exponents}
       \end{tabular*}
       {\rule{\temptablewidth}{1pt}}
       \end{center}
 \end{table}

 \subsection{Finite-size scaling in the equilibrium state}\label{subsec:FSSeq}
In the vicinity of the QCP with $N \rightarrow \infty$, one has
\begin{eqnarray*}
\xi &\sim& |q-q_c|^{-\nu}, \quad
 N_\xi \sim |q-q_c|^{-\nu d}\,,\label{eq:Len_Ninf}\\
 \Delta^{-1}&\sim& \xi^{z}\sim N_\xi^{z/d}\sim |q-q_c|^{-\nu z}\,,
 \end{eqnarray*}
 which shows the power-law divergence of the characteristic length and time at the critical point. At any finite $N$, the singularity at QCP thus gets rounded, the characteristic length $\xi$ would remain finite and a nonvanishing gap stays at the critical field. The ``rounding off'' can be introduced through a regular scaling function $g_\Delta(x)$, such that for the inverse gap
\begin{eqnarray}
\Delta^{-1}(q,N)& \sim & \Delta^{-1}(q,N=\infty)\cdot g_\Delta\left({N}/{N_\xi}\right)\,,\label{eq:gapFSS}
%\\&\sim&|q-q_c|^{-\nu z}\cdot g_\Delta(N/N_\xi)
\end{eqnarray}
with $g_\Delta(x)\rightarrow\text{const.}$ for $x\gg 1$, which recovers the nominal TDL, and $g_\Delta(x)\rightarrow x^{\omega_\Delta}$ for $x\ll 1$. The exponent $\omega_\Delta = z/d$ is obtained by assuming that $\Delta^{-1}$ would become regular at $q_c$ for any finite $N$. By using $z=1$ and $d=3$ obtained in last section, we find $\Delta\sim N^{-z/d}\sim N^{-1/3}$ at the pseudo-critical point because finite $N$ takes over the role of $N_\xi$ as a length scale cutoff. Such a scaling was already revealed  from fitting numerical calculated values in Ref.\,\cite{Zhang2013,Hoang2016a}. This is the same finite size behavior at the QCP as in the Dicke model \cite{jvidalDickeModel} and the Lipkin-Meshkov-Glick model \cite{Dusuel2004,Leyvraz2005}.

\par Based on the above discussions, the finite-size scaling hypotheses for the gap and order parameters can be generally chosen as,
\begin{eqnarray}
  \Delta(\epsilon, N) &\sim& N^{-z/d}g_1(\epsilon N^{1/\nu d})\,,\label{eq:Gapfss}\\
 \mathcal{O}(\epsilon,N) &\sim& N^{-\beta_\mathcal{O}/{\nu d}}g_\mathcal{O}(\epsilon N^{1/\nu d})\,,\label{eq:OPfss}
\end{eqnarray}
where $\epsilon = (q-q_c)/{q_c}$ is the reduced control parameter which measures the distance to QCP. The exponent $\beta_\mathcal{O}$ is the corresponding scaling dimension for observable $\mathcal{O}\,(\mathcal{O}=\mathcal{N},\mathcal{M})$, and $g_{1},\, g_\mathcal{O}$ are the scaling functions.
\par We numerically diagonalize the Hamiltonian in the $F_z = 0$ subspace for different size $N$ to obtain the gap $\Delta(q)=E_1(q)-E_0(q)$, ground state fractional population $\mathcal{N}(q)$ and transverse magnetization $\mathcal{M}(q)$. In Fig.\,\ref{fig:rescale_eq}, we show the data collapse by using mean-field critical exponents in Table \ref{tab:exponents}. The scaling hypotheses in Eqs.\,(\ref{eq:Gapfss})-(\ref{eq:OPfss}) are thus well verified near the QCP for the spin mixing model we discuss.

\begin{figure*}[!hbt]
  \centering
   \includegraphics[width=0.62\columnwidth]{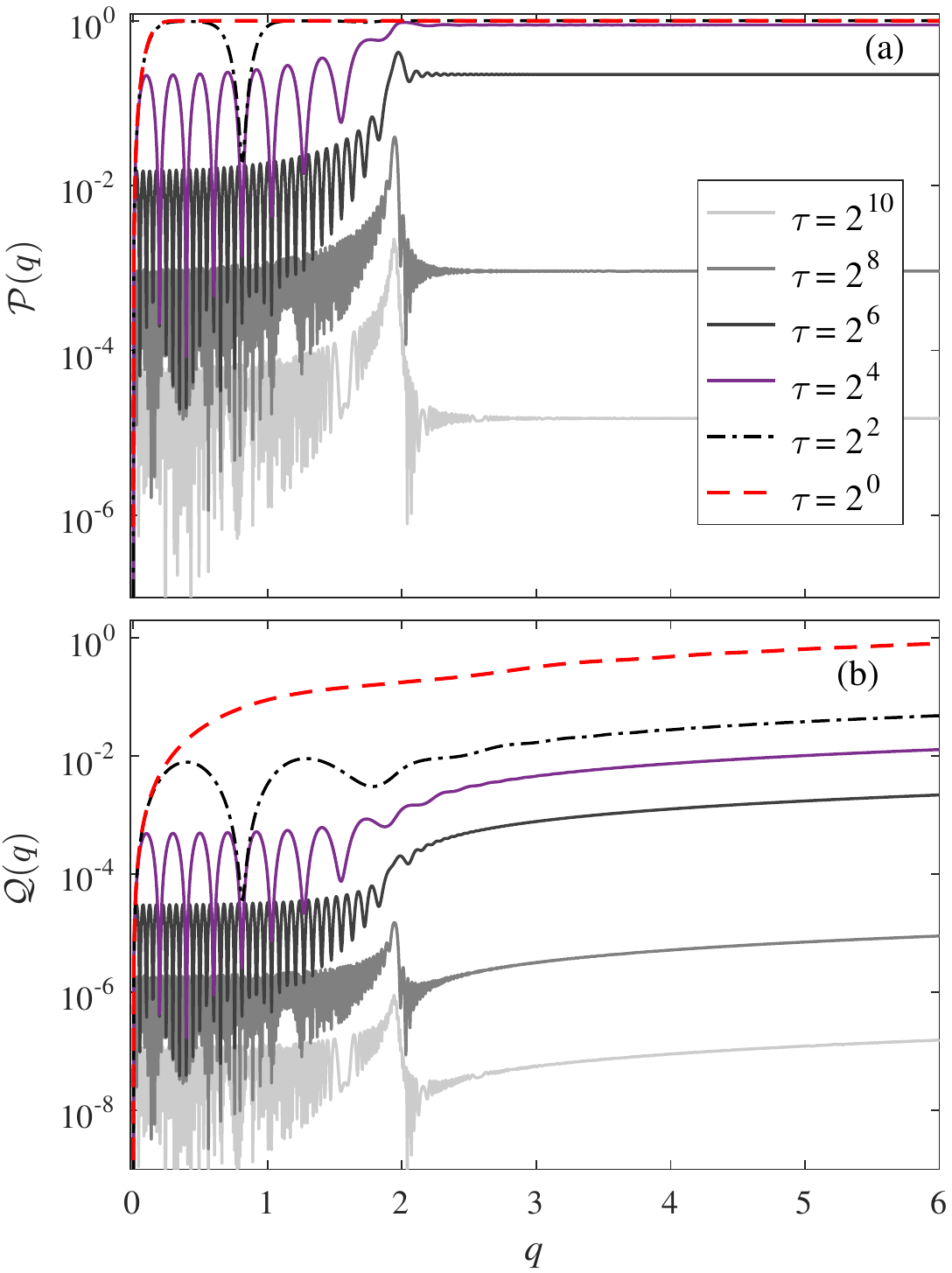}
   \quad
  \includegraphics[width=0.94\columnwidth]{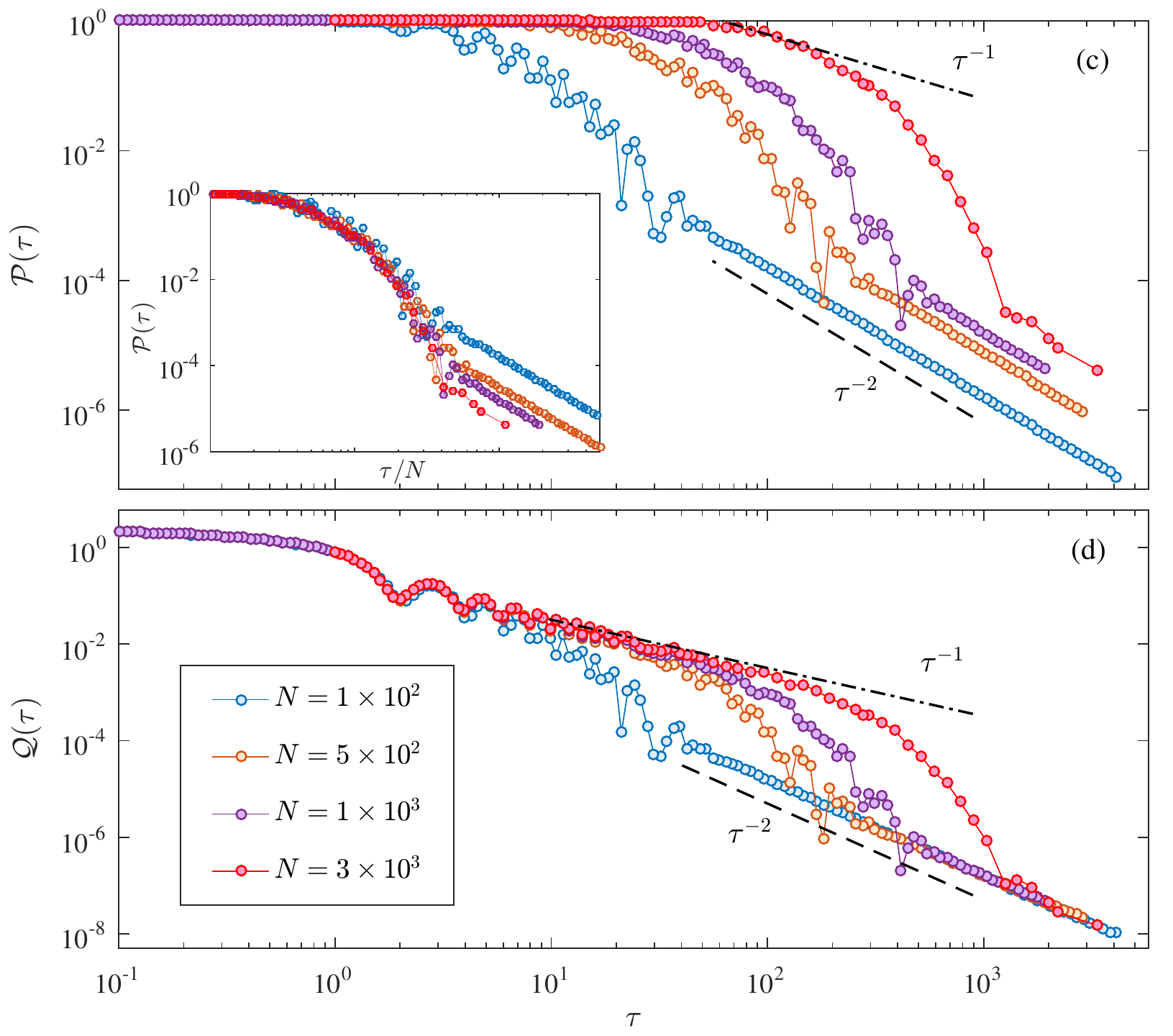}
  \caption{{\bf Driven dynamics.} (a) and (b) show the general structures of excitation probability $\mathcal{P}(q)$ and $\mathcal{Q}(q)$ at different driving rate, for $N = 1000$ as an example. (c)-(d) The excitation probability $P(\tau)$ and the heat density $\mathcal{Q}(\tau)$ at the end of the driving for different system size $N$. The driving parameters are taken as $q_i =0 $ and $ q_f = 6$. Three distinct dynamical regions are revealed according to the behaviors of $\mathcal{P}(\tau)$ and $\mathcal{Q}(\tau)$. The black dashed line and dash-dotted lines indicate the $\tau^{-2}$ and $\tau^{-1}$ power laws, respectively. Inset of (c), we rescale $\tau$-axis by $N$ and show the crossover between the adiabatic region and non-adiabatic region occurs at $\tau_c\propto N$ (see main text). }\label{fig:PexQ}
\end{figure*}

\section{Dynamic behaviors across the QCP}\label{sec:dynamic}
 The equilibrium criticality established above allows us to study the universal behaviors in the driven dynamics across the QCP. In this section, we discuss such behaviors for the driven dynamics in our model.
\par We consider the case of a linear driving protocol with the quadratic Zeeman shift in Eq.\,(\ref{eq:Hamil0}) taking the form,
 \begin{equation}
   q(t) = q_i + (q_f - q_i)\cdot t/\tau,\quad\text{for}\quad t\in[0,\tau] \,,\label{eq:protocol}
 \end{equation}
 where $q_i\equiv q(0),\, q_f\equiv q(\tau)$ are the initial and final shifts respectively, and $\tau$ is the total driving duration and driving speed is $v = \frac{q_f - q_i}{\tau}\propto\tau^{-1}$.  If $\tau\rightarrow 0$, such a driving protocol reduces to a sudden quench, while it corresponds to the adiabatic limit when $\tau\rightarrow \infty$. The initial state $|\Psi(t=0)\rangle$ is always taken to be the ground state of Hamiltonian $\hat H(q_i)$.
The dynamical state $|\Psi(t)\rangle$ is solved numerically by evolving the Schr\"odinger equation $i\partial_t|\Psi(t)\rangle = \hat H(t)|\Psi(t)\rangle\,$, with the driving protocol $\hat H(t)\equiv\hat H[q(t)]$ of Eq.\,(\ref{eq:protocol}).  Since only two parameters out of the three $(t,\, q,\, \tau)$ are independent, we can use either $(t,\tau)$ or $(q, \tau)$ to denote the same driving process in the following discussion, {\it i.e.}, $\mathcal{O}(q)\equiv \mathcal{O}[q(t)]$ for any time-dependent observables $\mathcal{O}$.
\par
One can always expand the state $|\Psi(q)\rangle$ as
$
  |\Psi(q)\rangle = \sum_{n=0}^{\mathcal{D}-1} a_n(q) e^{-i\Theta_n(q)} |\psi_n(q)\rangle \,,
$
into the instantaneous eigenstates $ |\psi_n(q)\rangle\,( n\in\mathbb{N})$ of $\hat H(q)$ satisfying $\hat H(q)|\psi_n(q)\rangle=E_n(q)|\psi_n(q)\rangle$. $\{a_n\}$ is the coefficients of superposition and $\mathcal{D}$ is the dimension of Hilbert space.
The time-dependent Schr\"odinger equation then reduces to
\begin{eqnarray*}
  \partial_t a_n(t)& = &- \sum_{m=0}^{\mathcal{D}-1} a_m(t)
  e^{i\left[\Theta_n(t)-\Theta_m(t)\right]}
  \langle\psi_n(t)|\partial_t|\psi_m(t)\rangle\,,
\end{eqnarray*}
  where the dynamical phase takes the explicit form $\Theta_n(q)=\int_{q_i}^q \frac{E_n(q^\prime)}{\dot{q}^\prime}dq^\prime=v\int_{q_i}^q {E_n(q^\prime)}dq^\prime$.
\par We characterize the loss of adiabaticity employing the following two quantities: the excitation probability $\mathcal{P}(t)=1-|\langle\Psi(t)|\psi_0(t)\rangle|^2$ which measures the infidelity of the dynamical state $|\Psi(t)\rangle$ on the adiabatically connected ground state $|\psi_0(t)\rangle$ and the excess heat density $\mathcal{Q}(t)=[\langle\Psi(t)|\hat H(t)|\Psi(t)\rangle-E_0(t)]/N$,
 which measures the overall net energy gain over $E_0(t)\equiv\langle\psi_0(t)|\hat H(t)|\psi_0(t)\rangle$. Starting from the ground state, with $\mathcal{P}(q_i)=0$ and $\mathcal{Q}(q_i)=0$,  we expect $1\geq \mathcal{P}(t)\geq 0$ and $ \mathcal{Q}(t)\geq 0$.
\par This study is focused on driving the system from BA phase ($q_i = 0$) to deep in the polar phase ($q_f=6$).
 When the system is driven across the QCP, due to the vanishing gap at the critical field, non-adiabatic effects become unavoidable even if the driving velocity $v\rightarrow 0$. For a finite-size system, the gap remains finite, and the dynamics show quite different behaviors in the limit $v\rightarrow 0$. This constitutes an important topic to be addressed in the following.
  \par Based on numerical simulations, we find there exist three distinct regions according to the driving rate and will be called adiabatic, non-adiabatic, and far-from-equilibrium region respectively corresponding to long, intermediate, and short $\tau$. Their non-adiabatic indicators show quite different scaling behaviors and are essentially decided by the dominant time or length scales and the corresponding low energy excitations in the driven processes.

{\it The adiabatic region for large $\tau$.}---For a large but finite $N$, a finite gap exists. Adiabatically passing through the pseudo-critical point is possible in the adiabatic perturbation limit $v\rightarrow 0$, when the system can only be excited by the so-called Landau-Zener mechanism. The adiabatic perturbation theory \cite{DeGrandi2010} gives
 \begin{widetext}
 \begin{eqnarray}
   |a_n(q)|^2 &\approx& v^2 \left\{ \left[\frac{|\langle\psi_n|\partial_{q_i}|\psi_0\rangle|^2}{(E_n(q_i)-E_0(q_i))^2} +\frac{|\langle\psi_n|\partial_{q}|\psi_0\rangle|^2}{(E_n(q)-E_0(q))^2}\right]
     -2
    \frac{\langle\psi_n|\partial_{q_i}|\psi_0\rangle}{E_n(q_i)-E_0(q_i)}
    \frac{\langle\psi_n|\partial_{q}|\psi_0\rangle}{E_n(q)-E_0(q)}\cos[\delta\Theta_{n0}]\right\}
    \,, \label{eq:apt}
 \end{eqnarray}
 \end{widetext}
 where the accumulated phase difference between the $n$-th excited state and the ground state is defined as $ \delta\Theta_{n0}=\Theta_n(q)-\Theta_0(q)=v\int_{q_i}^q [E_n(q^\prime)-E_0(q^\prime)]d q^\prime$. Provided that only the dominant excitation into the first excited state is considered, we find $\delta\Theta_{10}= v\int_{q_i}^q\Delta(q^\prime) dq^\prime$, see Fig.\,\ref{fig:rescale_eq}\,(a). The integration of the gap ensures $\delta\Theta_{10}(q)$ be a continuous and monotonous increasing function of $q$ and linearly depend on $v$. Therefore, the two terms in Eq.\,(\ref{eq:apt}) can well describe the amplitude and oscillation behaviors of $\mathcal{P}(q)\approx|a_1(q)|^2$ as shown in Fig.\,\ref{fig:PexQ}\,(a), respectively. For a specific large $\tau$, $\mathcal{P}(q)$ shows slow oscillations with large envelope around the QCP and fast oscillations with small envelope away from the QCP. This is due to the gap closing near the QCP, which leads to a slower growth of $\delta\Theta_{10}$. The linear dependence on driving rate $v$ for $\delta\Theta_{10}$ is revealed by the oscillation period structure, shown respectively in Figs.\,\ref{fig:PexQ}\,(a) and (b), reminiscent of a Russian doll collection, between protocols with different $v$.
 \par In this adiabatic region, diabatic effects induced by the external driving enter only as a perturbation near the QCP. It is clear that the final excitation probability $\mathcal{P} (\tau)$ and excess heat density $\mathcal{Q}(\tau)$ both show the $\sim v^2\propto\tau^{-2}$ scaling for a generic gapped system \cite{Polkovnikov2008a}, as predicted by Eq.\,(\ref{eq:apt}), and also visibly confirmed in the large $\tau$ region in Figs.\,\ref{fig:PexQ}\,(c)-(d). The finite energy gap $\Delta_\text{min}$ at the QCP is the dominant energy scale during the dynamics, or the finite size $N$ is the smallest and dominant length scale. One can thus define a size-dependent KZ rate as $v_{\rm KZ}(N)\sim N^{-{(1+\nu z)}/{\nu d}}$ or equivalently  a time scale $\tau_{\rm KZ}(N)\sim N^{{(1+\nu z)}/{\nu d}}$,  with such driving rate or time the correlation length $N_\xi$ at the frozen moment is of the order of the system size $N$. When $v$ is smaller than $v_{\rm KZ}(N)$, the system always remains adiabatic \cite{Huang2014}.

%%%%%%%%%%%%%%%%%%%%%%%%%%%%%%%%%%%%%%%%%%%%%%%%%%%%%%%%%%%%%%%%
{\it The non-adiabatic universal region.}---In this intermediate region, $v> v_{\rm KZ}(N)$ but remains much less than the relevant initial gap. The non-adiabatic indicators $\mathcal{P}(\tau)$ and $\mathcal{Q}(\tau)$ exhibit distinct behaviors from the adiabatic region. It is due to the existence of another external time\,(length) scale $t_{\rm KZ}\,(\xi_{\rm KZ})$ which dominates near the QCP. This so-called KZ time $t_{\rm KZ}\sim v^{-\nu z/(1+\nu z)}$ or KZ length scale $\xi_{\rm KZ}\sim v^{-\nu/(1+\nu z)}$ , is determined by the external driving, and acts as the smallest time or length scale in the universal dynamics near the QCP. The crossover between the two regions occurs when $v\simeq v_{\rm KZ}$, which predicts the crossover happens at $\tau_c\propto N$ for different system size $N$, as shown in the inset of Fig.\,\ref{fig:PexQ}\,(c). Analogously, we can define a maximal defect-free size $N_{\rm KZ}\sim \xi_{\rm KZ}^{d}\sim v^{-d\nu/(1+\nu z)}$ or an effective length scale given by the driving, and the defect density from the KZ mechanism is proportional to $1/N_{\rm KZ}$. Therefore we find $\mathcal{P}(\tau)\sim {1}/{N_{\rm KZ}}\sim v^{d\nu/(1+\nu z)}$ and $\mathcal{Q}(\tau)\sim\mathcal{P}(\tau)\sim v^{d\nu/(1+\nu z)}$ \cite{Kolodrubetz2012b,Kolodrubetz2015,dutta2015quantum}.
This KZ scaling is expected to hold in the limit of $v\rightarrow 0\; (\tau\rightarrow\infty)$ in the TDL [black dash-dot line in Figs.\,\ref{fig:PexQ}\,(c)-(d)]. The asymptotic behavior for $N\rightarrow\infty$ implies there adiabatic processes are excluded in the TDL. We recall the limits of $v\rightarrow 0$ ({\it i.e.}, $\tau\rightarrow\infty$) and  $N\rightarrow\infty$ do not commute \cite{Polkovnikov2008a}.

The above two regions respectively correspond to the adiabatic finite-size scaling (FSS) regime and the impulse finite-time scaling (FTS) regime of a finite-size system considered earlier in Ref.\,\cite{Huang2014}.
In the FSS regime, $N<N_{\xi}$ and $N<N_{\rm KZ}$, for example
 $\mathcal{P} = N^{-1}f_1(vN^{\frac{1+\nu z}{\nu d}})$
 and we have only considered the excitation at the QCP $\epsilon=0$. The argument $x=vN^{\frac{1+\nu z}{\nu d}}=vN$ is small and the scaling function $f_1(x)$ can be described perturbatively \cite{Huang2014,Liu2014} in $x$. Therefore we have $\mathcal{P}\simeq N^{-1}[f_1(0)+  f_1^\prime(0)\cdot x+\frac{1}{2}f_1^{\prime\prime}(0)\cdot x^2]$, where the first term $f_1(0)$  is the equilibrium excitation and should vanish for a finite system, the second and the third term arise from the perturbation of the driving and we argue that the linear term in $v$ is absent because the excitation or excess heat is insensitive to the sign of $v$ \cite{Polkovnikov2008a}, therefore we have $\mathcal{P}\simeq N^{-1}\cdot\frac{1}{2}f_1^{\prime\prime}(0)\cdot x^2\sim \tau^{-2}$.
\par
In a general scenario of KZ ramp, the tuning parameter is swept from the deep disordered phase (polar) to the ordered phase (BA). Due to the gap closing from $q=0$ to $q < 0$ and the appearance of a second QCP at $q=-2$, we choose to drive from the BA to the polar phase in order to obtain a steady value of $\mathcal{P}$ for a long ramp time.
In order to address the experiments, according to Ref.\,\cite{Gong2010,DeGrandi2011}, the order parameters easily measurable in experiments satisfy the dynamical KZ scaling form,
\begin{eqnarray}
 \mathcal{O}(\epsilon,v) &=&  v^{\frac{\beta_\mathcal{O}}{1+\nu z}}f_ \mathcal{O}(\epsilon v^{-\frac{1}{1+\nu z}}, Nv^{\frac{\nu d}{1+\nu z}})\label{eq:KZscaling}
\end{eqnarray}
where $\mathcal{O} = \langle\mathcal{\hat O}\rangle$ can be either $\mathcal{N}$ or $\mathcal{M}$, $\beta_\mathcal{O}$ is the corresponding critical exponents given in Table \ref{tab:exponents}.  $f_ \mathcal{O}(x,y)$ is a scaling function of arguments $(x,y)$, taking the FTS form in Ref.\,\cite{Huang2014} with finite-size effects included.

In actual experiments, one can easily prepare the initial state in the polar phase with all atoms in $|1,m_f=0\rangle$ state (the $F_z = 0$ subspace) and tune the quadratic Zeeman shift $q$ in Eq.\,(\ref{eq:Hamil0}) linearly as in Eq.\,(\ref{eq:protocol}) with different driving time $\tau$. During the tuning process, the dynamical values of the fractional population $\mathcal{N}$ and the transverse magnetization $\mathcal{M}$ can be measured in successive realizations. One can also vary the system size $N$ to take the finite-size scaling into consideration. The scaling hypothesis in Eq.\,(\ref{eq:KZscaling}) can be checked by doing data collapse in the two scaling directions with the experimental results.
\par
\begin{figure}%[!htb]
  \centering
 \includegraphics[width=1.0\columnwidth]{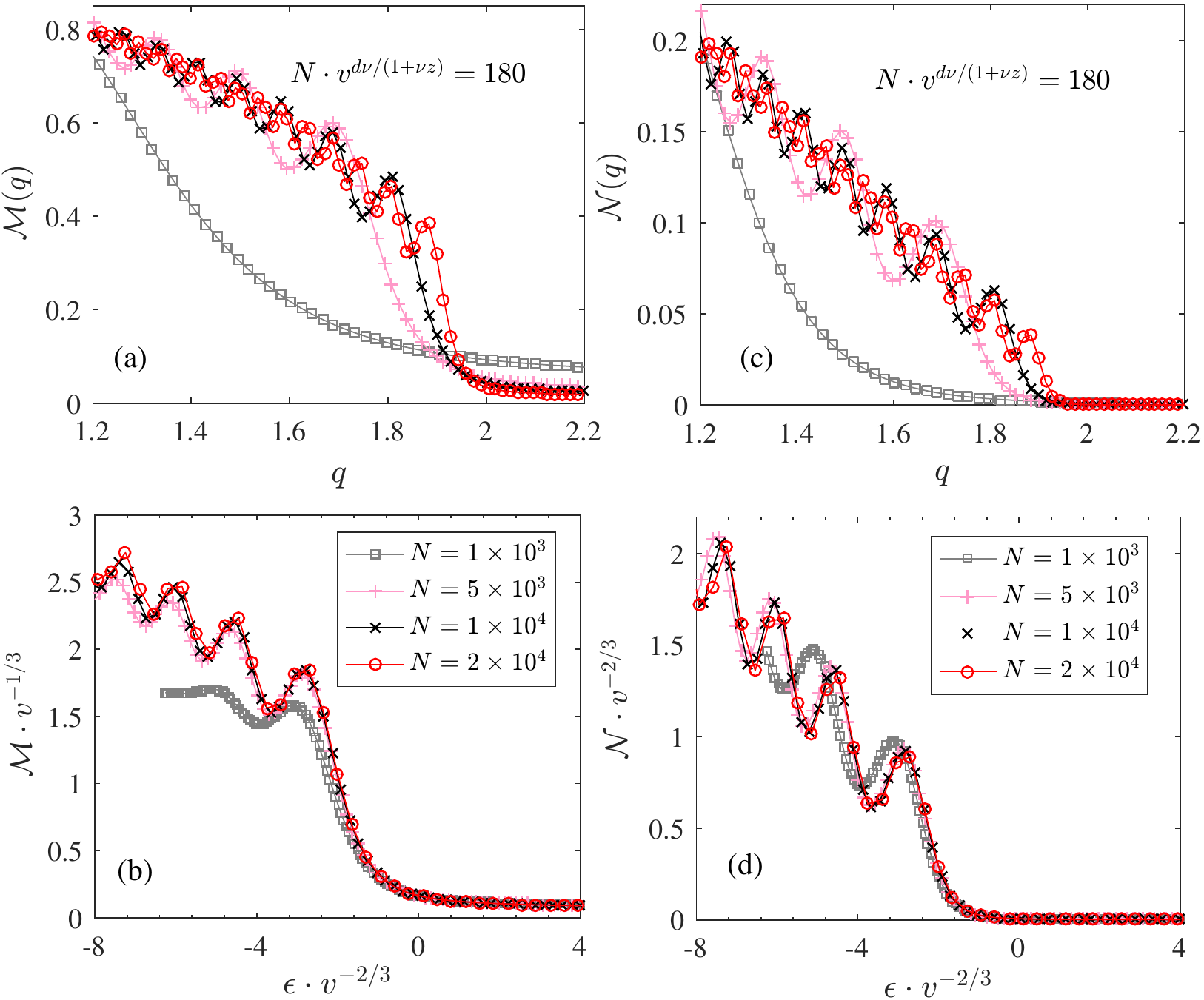}
  \caption{{\bf Finite-size Kibble-Zurek scaling.} For fixed $N\cdot v^{d\nu/(1+\nu z)}=N\cdot v = 180$ in Eq.\,(\ref{eq:KZscaling}) and starting from the polar phase ($q_i=4.0$) and sweeping to the BA phase $(q_f = 0)$.
  (a) The dynamical value of $\mathcal{N}(q)$ -- the fractional population. (b) The numerical data rescaled for $\mathcal{N}$.  (c) The transverse magnetization $\mathcal{M}(q)$. (d) The numerical data rescaled for $\mathcal{M}$.  For $N = 1\times 10^3, 5\times 10^3, 1\times 10^4 \text{ and } 2\times10^4$ which are all within experimentally feasible atom numbers. It is clear that in (b) and (d)  the KZ scaling hypotheses are verified near the QCP, but for smaller system size (gray line with square marker), the collapsed region shrinks. This indicates the loss of universality when $v$ is too fast.
  }\label{fig:dynamics}
\end{figure}

 We numerically check the full dynamical KZ scaling form by fixing $Nv^{\nu d/(1+\nu z)}=\text{const.}$, Fig.\,\ref{fig:dynamics}\,(a) and (c) show the numerically computed $\mathcal{M}$ and $\mathcal{N}$ with elected experimentally feasible system size $N$. These curves are indeed seen to collapse onto each other after rescaling according to Eq.\,(\ref{eq:KZscaling}), see Fig.\,\ref{fig:dynamics}\,(b) and (d). We note that for a small system size $N$, the scaling collapse region shrinks, which indicates the universality would disappear for the really small $\tau$ (large $v$) region.

{\it The far-from-equilibrium region for fast driving.}---
When the driving rate $v$ is too fast such that the driving determined length scale $N_{\rm KZ}$ is not only dominant near the QCP, but also during the whole dynamics as $N_{\rm KZ}< N_{\xi}(q_i)$, the system state becomes frozen in the whole driving process. The excitation probability $\mathcal{P}(q)$ saturates rapidly in the initial ramp and loses its feature as an indicator, as shown in Fig.\,\ref{fig:PexQ}\,(a) and (c). The heat density $\mathcal{Q}(\tau)$ shows almost no size dependence since the finite-size effects are unimportant at the initial gap $\Delta(q_i)$ [Fig.\,\ref{fig:rescale_eq}\,(a)], and $\mathcal{Q}(\tau)$ tends to nearly a constant for $\tau\rightarrow0$ as shown in Fig.\,\ref{fig:PexQ}\,(d). This far-from-equilibrium region by fast driving is non-universal.

\section{Discussions and Conclusions}\label{sec:conclusion}
In this paper, we study the equilibrium and dynamical properties in a ferromagnetic spinor atomic Bose-Einstein condensate. At equilibrium, we extract the mean-field critical exponents and verify the finite-size scaling hypothesis. Because of the infinitely long-range nature of the interaction (within the SMA), the mean-field theory gives exact results about the critical phenomena in the equilibrium.

The dynamical process is realized by linearly tuning the quadratic Zeeman shift across a continuous QCP.
In the vicinity of the QCP, universal behaviors are also observed in the dynamics. Three distinct dynamical regions are identified corresponding to different total driving time $\tau$\,(or equivalently driving rate $v\propto\tau^{-1}$), characterized by two adiabaticity indicators: the excitation probability $\mathcal{P}$ and the excess heat density $\mathcal{Q}\,$.
We show that the adiabatic region of $\,\mathcal{P}\sim\mathcal{Q}\sim\tau^{-2}\,$ exists in any finite system for $v<v_{\rm KZ}(N)$, in which external driving enters the dynamics only as a perturbation. In this region the adiabatic perturbation theory can give a nice description for the dynamics.  While the non-adiabatic universal region of $\,\mathcal{P}\sim\mathcal{Q}\sim\tau^{-\nu d/(1+\nu z)}\,$, which corresponds to intermediate driving rate $v>v_{\rm KZ}(N)$, and in the thermodynamic limit is well described by the Kibble-Zurek mechanism.
The dynamical Kibble-Zurek scaling is found to apply to finite-size systems in this universal region
and the scaling hypotheses for fractional population $\mathcal{N}$ and transverse magnetization $\mathcal{M}$ are presented which can be checked directly in ongoing experiments.
Finally, the region of the fastest driving rate is found to be non-universal and far-from-equilibrium with $\mathcal{P}$ and $\mathcal{Q}$ essentially being constants independent of $\tau$.

The distinct behaviors of the dynamics originate from the competitions between different length scales, the scale given by the external driving $N_{\rm KZ}$, the intrinsic correlation length scale of the system $N_{\xi}$, and the finite size $N$. The smallest one always dominates the dynamic behavior.
We also note that the above three regions: adiabatic, non-adiabatic and far-from-equilibrium regions may respectively correspond to the analytical, non-adiabatic, and non-analytical processes in Ref.\,\cite{Polkovnikov2008a}. As pointed by the authors of Ref.\,\cite{Polkovnikov2008a}, in the analytical and non-analytical regimes, there exist no highly populated low-energy modes and finite-size or relaxation effects are unimportant.
\par Finally, we emphasize that the simplicity and rich magnetic phases of spinor condensates could offer us a promising platform to study the critical phenomena theoretically and experimentally, both in equilibrium and the nonequilibrium.

{\it Note added.}---A related work addressing the similar topic but in the Lipkin-Meshkov-Glick model appeared in the archive very recently \cite{Defenu2018}.
\section*{Acknowledgement}
This work is supported by the National Basic Research Program of China (973 program) (No. 2013CB922004), NSFC (No. 11574100, No. 91636213 and No. 11747605).
S.Y. is supported in part by China Postdoctoral Science Foundation (Grant No. 2017M620035).

\bibliography{mybib}

%Control: production of eprint (-1) disabled
\begin{thebibliography}{64}%
\makeatletter
\providecommand \@ifxundefined [1]{%
 \@ifx{#1\undefined}
}%
\providecommand \@ifnum [1]{%
 \ifnum #1\expandafter \@firstoftwo
 \else \expandafter \@secondoftwo
 \fi
}%
\providecommand \@ifx [1]{%
 \ifx #1\expandafter \@firstoftwo
 \else \expandafter \@secondoftwo
 \fi
}%
\providecommand \natexlab [1]{#1}%
\providecommand \enquote  [1]{``#1''}%
\providecommand \bibnamefont  [1]{#1}%
\providecommand \bibfnamefont [1]{#1}%
\providecommand \citenamefont [1]{#1}%
\providecommand \href@noop [0]{\@secondoftwo}%
\providecommand \href [0]{\begingroup \@sanitize@url \@href}%
\providecommand \@href[1]{\@@startlink{#1}\@@href}%
\providecommand \@@href[1]{\endgroup#1\@@endlink}%
\providecommand \@sanitize@url [0]{\catcode `\\12\catcode `\$12\catcode
  `\&12\catcode `\#12\catcode `\^12\catcode `\_12\catcode `\%12\relax}%
\providecommand \@@startlink[1]{}%
\providecommand \@@endlink[0]{}%
\providecommand \url  [0]{\begingroup\@sanitize@url \@url }%
\providecommand \@url [1]{\endgroup\@href {#1}{\urlprefix }}%
\providecommand \urlprefix  [0]{URL }%
\providecommand \Eprint [0]{\href }%
\providecommand \doibase [0]{http://dx.doi.org/}%
\providecommand \selectlanguage [0]{\@gobble}%
\providecommand \bibinfo  [0]{\@secondoftwo}%
\providecommand \bibfield  [0]{\@secondoftwo}%
\providecommand \translation [1]{[#1]}%
\providecommand \BibitemOpen [0]{}%
\providecommand \bibitemStop [0]{}%
\providecommand \bibitemNoStop [0]{.\EOS\space}%
\providecommand \EOS [0]{\spacefactor3000\relax}%
\providecommand \BibitemShut  [1]{\csname bibitem#1\endcsname}%
\let\auto@bib@innerbib\@empty
%</preamble>
\bibitem [{\citenamefont {Sachdev}(2011)}]{sachdev2011quantum}%
  \BibitemOpen
  \bibfield  {author} {\bibinfo {author} {\bibfnamefont {S.}~\bibnamefont
  {Sachdev}},\ }\href@noop {} {\emph {\bibinfo {title} {Quantum Phase
  Transitions}}}\ (\bibinfo  {publisher} {Cambridge University Press},\
  \bibinfo {year} {2011})\BibitemShut {NoStop}%
\bibitem [{\citenamefont {Cardy}(1996)}]{cardy1996scaling}%
  \BibitemOpen
  \bibfield  {author} {\bibinfo {author} {\bibfnamefont {J.}~\bibnamefont
  {Cardy}},\ }\href@noop {} {\emph {\bibinfo {title} {Scaling and
  renormalization in statistical physics}}}\ (\bibinfo  {publisher} {Cambridge
  University Press},\ \bibinfo {year} {1996})\BibitemShut {NoStop}%
\bibitem [{\citenamefont {Stanley}(1999)}]{Stanley1999}%
  \BibitemOpen
  \bibfield  {author} {\bibinfo {author} {\bibfnamefont {H.~E.}\ \bibnamefont
  {Stanley}},\ }\href {\doibase 10.1103/RevModPhys.71.S358} {\bibfield
  {journal} {\bibinfo  {journal} {Rev. Mod. Phys.}\ }\textbf {\bibinfo {volume}
  {71}},\ \bibinfo {pages} {S358} (\bibinfo {year} {1999})}\BibitemShut
  {NoStop}%
\bibitem [{\citenamefont {Bloch}\ \emph {et~al.}(2008)\citenamefont {Bloch},
  \citenamefont {Dalibard},\ and\ \citenamefont {Zwerger}}]{Bloch2008}%
  \BibitemOpen
  \bibfield  {author} {\bibinfo {author} {\bibfnamefont {I.}~\bibnamefont
  {Bloch}}, \bibinfo {author} {\bibfnamefont {J.}~\bibnamefont {Dalibard}}, \
  and\ \bibinfo {author} {\bibfnamefont {W.}~\bibnamefont {Zwerger}},\ }\href
  {\doibase 10.1103/RevModPhys.80.885} {\bibfield  {journal} {\bibinfo
  {journal} {Rev. Mod. Phys.}\ }\textbf {\bibinfo {volume} {80}},\ \bibinfo
  {pages} {885} (\bibinfo {year} {2008})}\BibitemShut {NoStop}%
\bibitem [{\citenamefont {Polkovnikov}\ \emph {et~al.}(2011)\citenamefont
  {Polkovnikov}, \citenamefont {Sengupta}, \citenamefont {Silva},\ and\
  \citenamefont {Vengalattore}}]{Polkovnikov2011}%
  \BibitemOpen
  \bibfield  {author} {\bibinfo {author} {\bibfnamefont {A.}~\bibnamefont
  {Polkovnikov}}, \bibinfo {author} {\bibfnamefont {K.}~\bibnamefont
  {Sengupta}}, \bibinfo {author} {\bibfnamefont {A.}~\bibnamefont {Silva}}, \
  and\ \bibinfo {author} {\bibfnamefont {M.}~\bibnamefont {Vengalattore}},\
  }\href {\doibase 10.1103/RevModPhys.83.863} {\bibfield  {journal} {\bibinfo
  {journal} {Rev. Mod. Phys.}\ }\textbf {\bibinfo {volume} {83}},\ \bibinfo
  {pages} {863} (\bibinfo {year} {2011})}\BibitemShut {NoStop}%
\bibitem [{\citenamefont {Stamperkurn}\ and\ \citenamefont
  {Ueda}(2013)}]{stamperkurn2013spinor}%
  \BibitemOpen
  \bibfield  {author} {\bibinfo {author} {\bibfnamefont {D.~M.}\ \bibnamefont
  {Stamperkurn}}\ and\ \bibinfo {author} {\bibfnamefont {M.}~\bibnamefont
  {Ueda}},\ }\href
  {https://journals.aps.org/rmp/abstract/10.1103/RevModPhys.85.1191} {\bibfield
   {journal} {\bibinfo  {journal} {Rev. Mod. Phys.}\ }\textbf {\bibinfo
  {volume} {85}},\ \bibinfo {pages} {1191} (\bibinfo {year}
  {2013})}\BibitemShut {NoStop}%
\bibitem [{\citenamefont {Langen}\ \emph {et~al.}(2015)\citenamefont {Langen},
  \citenamefont {Geiger},\ and\ \citenamefont {Schmiedmayer}}]{Langen2015}%
  \BibitemOpen
  \bibfield  {author} {\bibinfo {author} {\bibfnamefont {T.}~\bibnamefont
  {Langen}}, \bibinfo {author} {\bibfnamefont {R.}~\bibnamefont {Geiger}}, \
  and\ \bibinfo {author} {\bibfnamefont {J.}~\bibnamefont {Schmiedmayer}},\
  }\href {\doibase 10.1146/annurev-conmatphys-031214-014548} {\bibfield
  {journal} {\bibinfo  {journal} {Annu. Rev. Condens. Matter Phys.}\ }\textbf
  {\bibinfo {volume} {6}},\ \bibinfo {pages} {201} (\bibinfo {year}
  {2015})}\BibitemShut {NoStop}%
\bibitem [{\citenamefont {Kibble}(1976)}]{kibble1976topology}%
  \BibitemOpen
  \bibfield  {author} {\bibinfo {author} {\bibfnamefont {T.~W.}\ \bibnamefont
  {Kibble}},\ }\href@noop {} {\bibfield  {journal} {\bibinfo  {journal}
  {Journal of Physics A: Mathematical and General}\ }\textbf {\bibinfo {volume}
  {9}},\ \bibinfo {pages} {1387} (\bibinfo {year} {1976})}\BibitemShut
  {NoStop}%
\bibitem [{\citenamefont {Zurek}(1985)}]{zurek1985cosmological}%
  \BibitemOpen
  \bibfield  {author} {\bibinfo {author} {\bibfnamefont {W.~H.}\ \bibnamefont
  {Zurek}},\ }\href@noop {} {\bibfield  {journal} {\bibinfo  {journal}
  {Nature}\ }\textbf {\bibinfo {volume} {317}},\ \bibinfo {pages} {505}
  (\bibinfo {year} {1985})}\BibitemShut {NoStop}%
\bibitem [{\citenamefont {Zurek}(1996)}]{zurek1996cosmological}%
  \BibitemOpen
  \bibfield  {author} {\bibinfo {author} {\bibfnamefont {W.~H.}\ \bibnamefont
  {Zurek}},\ }\href@noop {} {\bibfield  {journal} {\bibinfo  {journal} {Physics
  Reports}\ }\textbf {\bibinfo {volume} {276}},\ \bibinfo {pages} {177}
  (\bibinfo {year} {1996})}\BibitemShut {NoStop}%
\bibitem [{\citenamefont {Damski}(2005)}]{Damski2005}%
  \BibitemOpen
  \bibfield  {author} {\bibinfo {author} {\bibfnamefont {B.}~\bibnamefont
  {Damski}},\ }\href {\doibase 10.1103/PhysRevLett.95.035701} {\bibfield
  {journal} {\bibinfo  {journal} {Phys. Rev. Lett.}\ }\textbf {\bibinfo
  {volume} {95}},\ \bibinfo {pages} {035701} (\bibinfo {year}
  {2005})}\BibitemShut {NoStop}%
\bibitem [{\citenamefont {Zurek}\ \emph {et~al.}(2005)\citenamefont {Zurek},
  \citenamefont {Dorner},\ and\ \citenamefont {Zoller}}]{Zurek2005}%
  \BibitemOpen
  \bibfield  {author} {\bibinfo {author} {\bibfnamefont {W.~H.}\ \bibnamefont
  {Zurek}}, \bibinfo {author} {\bibfnamefont {U.}~\bibnamefont {Dorner}}, \
  and\ \bibinfo {author} {\bibfnamefont {P.}~\bibnamefont {Zoller}},\ }\href
  {\doibase 10.1103/PhysRevLett.95.105701} {\bibfield  {journal} {\bibinfo
  {journal} {Phys. Rev. Lett.}\ }\textbf {\bibinfo {volume} {95}},\ \bibinfo
  {pages} {105701} (\bibinfo {year} {2005})}\BibitemShut {NoStop}%
\bibitem [{\citenamefont {Damski}\ and\ \citenamefont
  {Zurek}(2006)}]{Damski2006}%
  \BibitemOpen
  \bibfield  {author} {\bibinfo {author} {\bibfnamefont {B.}~\bibnamefont
  {Damski}}\ and\ \bibinfo {author} {\bibfnamefont {W.~H.}\ \bibnamefont
  {Zurek}},\ }\href {\doibase 10.1103/PhysRevA.73.063405} {\bibfield  {journal}
  {\bibinfo  {journal} {Phys. Rev. A}\ }\textbf {\bibinfo {volume} {73}},\
  \bibinfo {pages} {063405} (\bibinfo {year} {2006})}\BibitemShut {NoStop}%
\bibitem [{\citenamefont {Damski}\ and\ \citenamefont
  {Zurek}(2007)}]{Damski2007}%
  \BibitemOpen
  \bibfield  {author} {\bibinfo {author} {\bibfnamefont {B.}~\bibnamefont
  {Damski}}\ and\ \bibinfo {author} {\bibfnamefont {W.~H.}\ \bibnamefont
  {Zurek}},\ }\href {\doibase 10.1103/PhysRevLett.99.130402} {\bibfield
  {journal} {\bibinfo  {journal} {Phys. Rev. Lett.}\ }\textbf {\bibinfo
  {volume} {99}},\ \bibinfo {pages} {130402} (\bibinfo {year}
  {2007})}\BibitemShut {NoStop}%
\bibitem [{\citenamefont {Lamacraft}(2007)}]{Lamacraft2007}%
  \BibitemOpen
  \bibfield  {author} {\bibinfo {author} {\bibfnamefont {A.}~\bibnamefont
  {Lamacraft}},\ }\href {\doibase 10.1103/PhysRevLett.98.160404} {\bibfield
  {journal} {\bibinfo  {journal} {Phys. Rev. Lett.}\ }\textbf {\bibinfo
  {volume} {98}},\ \bibinfo {pages} {160404} (\bibinfo {year}
  {2007})}\BibitemShut {NoStop}%
\bibitem [{\citenamefont {Saito}\ \emph {et~al.}(2007)\citenamefont {Saito},
  \citenamefont {Kawaguchi},\ and\ \citenamefont {Ueda}}]{Saito2007}%
  \BibitemOpen
  \bibfield  {author} {\bibinfo {author} {\bibfnamefont {H.}~\bibnamefont
  {Saito}}, \bibinfo {author} {\bibfnamefont {Y.}~\bibnamefont {Kawaguchi}}, \
  and\ \bibinfo {author} {\bibfnamefont {M.}~\bibnamefont {Ueda}},\ }\href
  {\doibase 10.1103/PhysRevA.76.043613} {\bibfield  {journal} {\bibinfo
  {journal} {Phys. Rev. A}\ }\textbf {\bibinfo {volume} {76}},\ \bibinfo
  {pages} {043613} (\bibinfo {year} {2007})}\BibitemShut {NoStop}%
\bibitem [{\citenamefont {Uhlmann}\ \emph {et~al.}(2007)\citenamefont
  {Uhlmann}, \citenamefont {Sch\"utzhold},\ and\ \citenamefont
  {Fischer}}]{Uwe2007}%
  \BibitemOpen
  \bibfield  {author} {\bibinfo {author} {\bibfnamefont {M.}~\bibnamefont
  {Uhlmann}}, \bibinfo {author} {\bibfnamefont {R.}~\bibnamefont
  {Sch\"utzhold}}, \ and\ \bibinfo {author} {\bibfnamefont {U.~R.}\
  \bibnamefont {Fischer}},\ }\href {\doibase 10.1103/PhysRevLett.99.120407}
  {\bibfield  {journal} {\bibinfo  {journal} {Phys. Rev. Lett.}\ }\textbf
  {\bibinfo {volume} {99}},\ \bibinfo {pages} {120407} (\bibinfo {year}
  {2007})}\BibitemShut {NoStop}%
\bibitem [{\citenamefont {Cucchietti}\ \emph {et~al.}(2007)\citenamefont
  {Cucchietti}, \citenamefont {Damski}, \citenamefont {Dziarmaga},\ and\
  \citenamefont {Zurek}}]{Cucchietti2007}%
  \BibitemOpen
  \bibfield  {author} {\bibinfo {author} {\bibfnamefont {F.~M.}\ \bibnamefont
  {Cucchietti}}, \bibinfo {author} {\bibfnamefont {B.}~\bibnamefont {Damski}},
  \bibinfo {author} {\bibfnamefont {J.}~\bibnamefont {Dziarmaga}}, \ and\
  \bibinfo {author} {\bibfnamefont {W.~H.}\ \bibnamefont {Zurek}},\ }\href
  {\doibase 10.1103/PhysRevA.75.023603} {\bibfield  {journal} {\bibinfo
  {journal} {Phys. Rev. A}\ }\textbf {\bibinfo {volume} {75}},\ \bibinfo
  {pages} {023603} (\bibinfo {year} {2007})}\BibitemShut {NoStop}%
\bibitem [{\citenamefont {del Campo}\ \emph {et~al.}(2010)\citenamefont {del
  Campo}, \citenamefont {{De Chiara}}, \citenamefont {Morigi}, \citenamefont
  {Plenio},\ and\ \citenamefont {Retzker}}]{DelCampo2010}%
  \BibitemOpen
  \bibfield  {author} {\bibinfo {author} {\bibfnamefont {A.}~\bibnamefont {del
  Campo}}, \bibinfo {author} {\bibfnamefont {G.}~\bibnamefont {{De Chiara}}},
  \bibinfo {author} {\bibfnamefont {G.}~\bibnamefont {Morigi}}, \bibinfo
  {author} {\bibfnamefont {M.~B.}\ \bibnamefont {Plenio}}, \ and\ \bibinfo
  {author} {\bibfnamefont {A.}~\bibnamefont {Retzker}},\ }\href {\doibase
  10.1103/PhysRevLett.105.075701} {\bibfield  {journal} {\bibinfo  {journal}
  {Phys. Rev. Lett.}\ }\textbf {\bibinfo {volume} {105}},\ \bibinfo {pages}
  {075701} (\bibinfo {year} {2010})}\BibitemShut {NoStop}%
\bibitem [{\citenamefont {Sabbatini}\ \emph {et~al.}(2011)\citenamefont
  {Sabbatini}, \citenamefont {Zurek},\ and\ \citenamefont
  {Davis}}]{Sabbatini2011}%
  \BibitemOpen
  \bibfield  {author} {\bibinfo {author} {\bibfnamefont {J.}~\bibnamefont
  {Sabbatini}}, \bibinfo {author} {\bibfnamefont {W.~H.}\ \bibnamefont
  {Zurek}}, \ and\ \bibinfo {author} {\bibfnamefont {M.~J.}\ \bibnamefont
  {Davis}},\ }\href {\doibase 10.1103/PhysRevLett.107.230402} {\bibfield
  {journal} {\bibinfo  {journal} {Phys. Rev. Lett.}\ }\textbf {\bibinfo
  {volume} {107}},\ \bibinfo {pages} {230402} (\bibinfo {year}
  {2011})}\BibitemShut {NoStop}%
\bibitem [{\citenamefont {Saito}\ \emph {et~al.}(2013)\citenamefont {Saito},
  \citenamefont {Kawaguchi},\ and\ \citenamefont {Ueda}}]{Saito2013a}%
  \BibitemOpen
  \bibfield  {author} {\bibinfo {author} {\bibfnamefont {H.}~\bibnamefont
  {Saito}}, \bibinfo {author} {\bibfnamefont {Y.}~\bibnamefont {Kawaguchi}}, \
  and\ \bibinfo {author} {\bibfnamefont {M.}~\bibnamefont {Ueda}},\ }\href
  {\doibase 10.1088/0953-8984/25/40/404212} {\bibfield  {journal} {\bibinfo
  {journal} {J. Phys. Condens. Matter}\ }\textbf {\bibinfo {volume} {25}},\
  \bibinfo {pages} {404212} (\bibinfo {year} {2013})}\BibitemShut {NoStop}%
\bibitem [{\citenamefont {Huang}\ \emph {et~al.}(2014)\citenamefont {Huang},
  \citenamefont {Yin}, \citenamefont {Feng},\ and\ \citenamefont
  {Zhong}}]{Huang2014}%
  \BibitemOpen
  \bibfield  {author} {\bibinfo {author} {\bibfnamefont {Y.}~\bibnamefont
  {Huang}}, \bibinfo {author} {\bibfnamefont {S.}~\bibnamefont {Yin}}, \bibinfo
  {author} {\bibfnamefont {B.}~\bibnamefont {Feng}}, \ and\ \bibinfo {author}
  {\bibfnamefont {F.}~\bibnamefont {Zhong}},\ }\href {\doibase
  10.1103/PhysRevB.90.134108} {\bibfield  {journal} {\bibinfo  {journal} {Phys.
  Rev. B}\ }\textbf {\bibinfo {volume} {90}},\ \bibinfo {pages} {134108}
  (\bibinfo {year} {2014})}\BibitemShut {NoStop}%
\bibitem [{\citenamefont {Lee}\ \emph {et~al.}(2015)\citenamefont {Lee},
  \citenamefont {Han},\ and\ \citenamefont {Choi}}]{Lee2015}%
  \BibitemOpen
  \bibfield  {author} {\bibinfo {author} {\bibfnamefont {M.}~\bibnamefont
  {Lee}}, \bibinfo {author} {\bibfnamefont {S.}~\bibnamefont {Han}}, \ and\
  \bibinfo {author} {\bibfnamefont {M.-S.}\ \bibnamefont {Choi}},\ }\href
  {\doibase 10.1103/PhysRevB.92.035117} {\bibfield  {journal} {\bibinfo
  {journal} {Phys. Rev. B}\ }\textbf {\bibinfo {volume} {92}},\ \bibinfo
  {pages} {035117} (\bibinfo {year} {2015})}\BibitemShut {NoStop}%
\bibitem [{\citenamefont {Jaschke}\ \emph {et~al.}(2017)\citenamefont
  {Jaschke}, \citenamefont {Maeda}, \citenamefont {Whalen}, \citenamefont
  {Wall},\ and\ \citenamefont {Carr}}]{Jaschke2016}%
  \BibitemOpen
  \bibfield  {author} {\bibinfo {author} {\bibfnamefont {D.}~\bibnamefont
  {Jaschke}}, \bibinfo {author} {\bibfnamefont {K.}~\bibnamefont {Maeda}},
  \bibinfo {author} {\bibfnamefont {J.~D.}\ \bibnamefont {Whalen}}, \bibinfo
  {author} {\bibfnamefont {M.~L.}\ \bibnamefont {Wall}}, \ and\ \bibinfo
  {author} {\bibfnamefont {L.~D.}\ \bibnamefont {Carr}},\ }\href {\doibase
  10.1088/1367-2630/aa65bc} {\bibfield  {journal} {\bibinfo  {journal} {New J.
  Phys.}\ }\textbf {\bibinfo {volume} {19}},\ \bibinfo {pages} {033032}
  (\bibinfo {year} {2017})}\BibitemShut {NoStop}%
\bibitem [{\citenamefont {Chuang}\ \emph {et~al.}(1991)\citenamefont {Chuang},
  \citenamefont {Durrer}, \citenamefont {Turok},\ and\ \citenamefont
  {Yurke}}]{chuang1991cosmology}%
  \BibitemOpen
  \bibfield  {author} {\bibinfo {author} {\bibfnamefont {I.}~\bibnamefont
  {Chuang}}, \bibinfo {author} {\bibfnamefont {R.}~\bibnamefont {Durrer}},
  \bibinfo {author} {\bibfnamefont {N.}~\bibnamefont {Turok}}, \ and\ \bibinfo
  {author} {\bibfnamefont {B.}~\bibnamefont {Yurke}},\ }\href@noop {}
  {\bibfield  {journal} {\bibinfo  {journal} {Science}\ }\textbf {\bibinfo
  {volume} {251}},\ \bibinfo {pages} {1336} (\bibinfo {year}
  {1991})}\BibitemShut {NoStop}%
\bibitem [{\citenamefont {B{\"a}uerle}\ \emph {et~al.}(1996)\citenamefont
  {B{\"a}uerle}, \citenamefont {Bunkov}, \citenamefont {Fisher}, \citenamefont
  {Godfrin},\ and\ \citenamefont {Pickett}}]{bauerle1996laboratory}%
  \BibitemOpen
  \bibfield  {author} {\bibinfo {author} {\bibfnamefont {C.}~\bibnamefont
  {B{\"a}uerle}}, \bibinfo {author} {\bibfnamefont {Y.~M.}\ \bibnamefont
  {Bunkov}}, \bibinfo {author} {\bibfnamefont {S.}~\bibnamefont {Fisher}},
  \bibinfo {author} {\bibfnamefont {H.}~\bibnamefont {Godfrin}}, \ and\
  \bibinfo {author} {\bibfnamefont {G.}~\bibnamefont {Pickett}},\ }\href@noop
  {} {\bibfield  {journal} {\bibinfo  {journal} {Nature}\ }\textbf {\bibinfo
  {volume} {382}},\ \bibinfo {pages} {332} (\bibinfo {year}
  {1996})}\BibitemShut {NoStop}%
\bibitem [{\citenamefont {Ruutu}\ \emph {et~al.}(1996)\citenamefont {Ruutu},
  \citenamefont {Eltsov}, \citenamefont {Gill}, \citenamefont {Kibble},
  \citenamefont {Krusius}, \citenamefont {Makhlin}, \citenamefont {Placais},
  \citenamefont {Volovik},\ and\ \citenamefont {Xu}}]{ruutu1996vortex}%
  \BibitemOpen
  \bibfield  {author} {\bibinfo {author} {\bibfnamefont {V.}~\bibnamefont
  {Ruutu}}, \bibinfo {author} {\bibfnamefont {V.}~\bibnamefont {Eltsov}},
  \bibinfo {author} {\bibfnamefont {A.}~\bibnamefont {Gill}}, \bibinfo {author}
  {\bibfnamefont {T.}~\bibnamefont {Kibble}}, \bibinfo {author} {\bibfnamefont
  {M.}~\bibnamefont {Krusius}}, \bibinfo {author} {\bibfnamefont {Y.~G.}\
  \bibnamefont {Makhlin}}, \bibinfo {author} {\bibfnamefont {B.}~\bibnamefont
  {Placais}}, \bibinfo {author} {\bibfnamefont {G.}~\bibnamefont {Volovik}}, \
  and\ \bibinfo {author} {\bibfnamefont {W.}~\bibnamefont {Xu}},\ }\href@noop
  {} {\bibfield  {journal} {\bibinfo  {journal} {Nature}\ }\textbf {\bibinfo
  {volume} {382}},\ \bibinfo {pages} {334} (\bibinfo {year}
  {1996})}\BibitemShut {NoStop}%
\bibitem [{\citenamefont {Chen}\ \emph {et~al.}(2011)\citenamefont {Chen},
  \citenamefont {White}, \citenamefont {Borries},\ and\ \citenamefont
  {DeMarco}}]{Chen2011}%
  \BibitemOpen
  \bibfield  {author} {\bibinfo {author} {\bibfnamefont {D.}~\bibnamefont
  {Chen}}, \bibinfo {author} {\bibfnamefont {M.}~\bibnamefont {White}},
  \bibinfo {author} {\bibfnamefont {C.}~\bibnamefont {Borries}}, \ and\
  \bibinfo {author} {\bibfnamefont {B.}~\bibnamefont {DeMarco}},\ }\href
  {\doibase 10.1103/PhysRevLett.106.235304} {\bibfield  {journal} {\bibinfo
  {journal} {Phys. Rev. Lett.}\ }\textbf {\bibinfo {volume} {106}},\ \bibinfo
  {pages} {235304} (\bibinfo {year} {2011})}\BibitemShut {NoStop}%
\bibitem [{\citenamefont {Baumann}\ \emph {et~al.}(2011)\citenamefont
  {Baumann}, \citenamefont {Mottl}, \citenamefont {Brennecke},\ and\
  \citenamefont {Esslinger}}]{Baumann2011}%
  \BibitemOpen
  \bibfield  {author} {\bibinfo {author} {\bibfnamefont {K.}~\bibnamefont
  {Baumann}}, \bibinfo {author} {\bibfnamefont {R.}~\bibnamefont {Mottl}},
  \bibinfo {author} {\bibfnamefont {F.}~\bibnamefont {Brennecke}}, \ and\
  \bibinfo {author} {\bibfnamefont {T.}~\bibnamefont {Esslinger}},\ }\href
  {\doibase 10.1103/PhysRevLett.107.140402} {\bibfield  {journal} {\bibinfo
  {journal} {Phys. Rev. Lett.}\ }\textbf {\bibinfo {volume} {107}},\ \bibinfo
  {pages} {140402} (\bibinfo {year} {2011})}\BibitemShut {NoStop}%
\bibitem [{\citenamefont {Lamporesi}\ \emph {et~al.}(2013)\citenamefont
  {Lamporesi}, \citenamefont {Donadello}, \citenamefont {Serafini},
  \citenamefont {Dalfovo},\ and\ \citenamefont {Ferrari}}]{Lamporesi2013}%
  \BibitemOpen
  \bibfield  {author} {\bibinfo {author} {\bibfnamefont {G.}~\bibnamefont
  {Lamporesi}}, \bibinfo {author} {\bibfnamefont {S.}~\bibnamefont
  {Donadello}}, \bibinfo {author} {\bibfnamefont {S.}~\bibnamefont {Serafini}},
  \bibinfo {author} {\bibfnamefont {F.}~\bibnamefont {Dalfovo}}, \ and\
  \bibinfo {author} {\bibfnamefont {G.}~\bibnamefont {Ferrari}},\ }\href
  {\doibase 10.1038/nphys2734} {\bibfield  {journal} {\bibinfo  {journal} {Nat
  Phys}\ }\textbf {\bibinfo {volume} {9}},\ \bibinfo {pages} {656} (\bibinfo
  {year} {2013})}\BibitemShut {NoStop}%
\bibitem [{\citenamefont {Corman}\ \emph {et~al.}(2014)\citenamefont {Corman},
  \citenamefont {Chomaz}, \citenamefont {Bienaim{\'{e}}}, \citenamefont
  {Desbuquois}, \citenamefont {Weitenberg}, \citenamefont {Nascimb{\`{e}}ne},
  \citenamefont {Dalibard},\ and\ \citenamefont {Beugnon}}]{Corman2014}%
  \BibitemOpen
  \bibfield  {author} {\bibinfo {author} {\bibfnamefont {L.}~\bibnamefont
  {Corman}}, \bibinfo {author} {\bibfnamefont {L.}~\bibnamefont {Chomaz}},
  \bibinfo {author} {\bibfnamefont {T.}~\bibnamefont {Bienaim{\'{e}}}},
  \bibinfo {author} {\bibfnamefont {R.}~\bibnamefont {Desbuquois}}, \bibinfo
  {author} {\bibfnamefont {C.}~\bibnamefont {Weitenberg}}, \bibinfo {author}
  {\bibfnamefont {S.}~\bibnamefont {Nascimb{\`{e}}ne}}, \bibinfo {author}
  {\bibfnamefont {J.}~\bibnamefont {Dalibard}}, \ and\ \bibinfo {author}
  {\bibfnamefont {J.}~\bibnamefont {Beugnon}},\ }\href {\doibase
  10.1103/PhysRevLett.113.135302} {\bibfield  {journal} {\bibinfo  {journal}
  {Phys. Rev. Lett.}\ }\textbf {\bibinfo {volume} {113}},\ \bibinfo {pages}
  {135302} (\bibinfo {year} {2014})}\BibitemShut {NoStop}%
\bibitem [{\citenamefont {Navon}\ \emph {et~al.}(2015)\citenamefont {Navon},
  \citenamefont {Gaunt}, \citenamefont {Smith},\ and\ \citenamefont
  {Hadzibabic}}]{Navon2015}%
  \BibitemOpen
  \bibfield  {author} {\bibinfo {author} {\bibfnamefont {N.}~\bibnamefont
  {Navon}}, \bibinfo {author} {\bibfnamefont {A.~L.}\ \bibnamefont {Gaunt}},
  \bibinfo {author} {\bibfnamefont {R.~P.}\ \bibnamefont {Smith}}, \ and\
  \bibinfo {author} {\bibfnamefont {Z.}~\bibnamefont {Hadzibabic}},\ }\href
  {\doibase 10.1126/science.1258676} {\bibfield  {journal} {\bibinfo  {journal}
  {Science}\ }\textbf {\bibinfo {volume} {347}},\ \bibinfo {pages} {167}
  (\bibinfo {year} {2015})}\BibitemShut {NoStop}%
\bibitem [{\citenamefont {Clark}\ \emph {et~al.}(2016)\citenamefont {Clark},
  \citenamefont {Feng},\ and\ \citenamefont {Chin}}]{Clark2016}%
  \BibitemOpen
  \bibfield  {author} {\bibinfo {author} {\bibfnamefont {L.~W.}\ \bibnamefont
  {Clark}}, \bibinfo {author} {\bibfnamefont {L.}~\bibnamefont {Feng}}, \ and\
  \bibinfo {author} {\bibfnamefont {C.}~\bibnamefont {Chin}},\ }\href {\doibase
  10.1126/science.aaf9657} {\bibfield  {journal} {\bibinfo  {journal}
  {Science}\ }\textbf {\bibinfo {volume} {354}},\ \bibinfo {pages} {606}
  (\bibinfo {year} {2016})}\BibitemShut {NoStop}%
\bibitem [{\citenamefont {Anquez}\ \emph {et~al.}(2016)\citenamefont {Anquez},
  \citenamefont {Robbins}, \citenamefont {Bharath}, \citenamefont
  {Boguslawski}, \citenamefont {Hoang},\ and\ \citenamefont
  {Chapman}}]{Anquez2016}%
  \BibitemOpen
  \bibfield  {author} {\bibinfo {author} {\bibfnamefont {M.}~\bibnamefont
  {Anquez}}, \bibinfo {author} {\bibfnamefont {B.~A.}\ \bibnamefont {Robbins}},
  \bibinfo {author} {\bibfnamefont {H.~M.}\ \bibnamefont {Bharath}}, \bibinfo
  {author} {\bibfnamefont {M.}~\bibnamefont {Boguslawski}}, \bibinfo {author}
  {\bibfnamefont {T.~M.}\ \bibnamefont {Hoang}}, \ and\ \bibinfo {author}
  {\bibfnamefont {M.~S.}\ \bibnamefont {Chapman}},\ }\href {\doibase
  10.1103/PhysRevLett.116.155301} {\bibfield  {journal} {\bibinfo  {journal}
  {Phys. Rev. Lett.}\ }\textbf {\bibinfo {volume} {116}},\ \bibinfo {pages}
  {155301} (\bibinfo {year} {2016})}\BibitemShut {NoStop}%
\bibitem [{\citenamefont {Aidelsburger}\ \emph {et~al.}(2017)\citenamefont
  {Aidelsburger}, \citenamefont {Ville}, \citenamefont {Saint-Jalm},
  \citenamefont {Nascimb{\`{e}}ne}, \citenamefont {Dalibard},\ and\
  \citenamefont {Beugnon}}]{Aidelsburger2017a}%
  \BibitemOpen
  \bibfield  {author} {\bibinfo {author} {\bibfnamefont {M.}~\bibnamefont
  {Aidelsburger}}, \bibinfo {author} {\bibfnamefont {J.~L.}\ \bibnamefont
  {Ville}}, \bibinfo {author} {\bibfnamefont {R.}~\bibnamefont {Saint-Jalm}},
  \bibinfo {author} {\bibfnamefont {S.}~\bibnamefont {Nascimb{\`{e}}ne}},
  \bibinfo {author} {\bibfnamefont {J.}~\bibnamefont {Dalibard}}, \ and\
  \bibinfo {author} {\bibfnamefont {J.}~\bibnamefont {Beugnon}},\ }\href
  {\doibase 10.1103/PhysRevLett.119.190403} {\bibfield  {journal} {\bibinfo
  {journal} {Phys. Rev. Lett.}\ }\textbf {\bibinfo {volume} {119}},\ \bibinfo
  {pages} {190403} (\bibinfo {year} {2017})}\BibitemShut {NoStop}%
\bibitem [{\citenamefont {Stenger}\ \emph {et~al.}(1998)\citenamefont
  {Stenger}, \citenamefont {Inouye}, \citenamefont {Stamper-Kurn},
  \citenamefont {Miesner}, \citenamefont {Chikkatur},\ and\ \citenamefont
  {Ketterle}}]{Stenger1998}%
  \BibitemOpen
  \bibfield  {author} {\bibinfo {author} {\bibfnamefont {J.}~\bibnamefont
  {Stenger}}, \bibinfo {author} {\bibfnamefont {S.}~\bibnamefont {Inouye}},
  \bibinfo {author} {\bibfnamefont {D.~M.~M.}\ \bibnamefont {Stamper-Kurn}},
  \bibinfo {author} {\bibfnamefont {H.-J.}\ \bibnamefont {Miesner}}, \bibinfo
  {author} {\bibfnamefont {A.~P.}\ \bibnamefont {Chikkatur}}, \ and\ \bibinfo
  {author} {\bibfnamefont {W.}~\bibnamefont {Ketterle}},\ }\href {\doibase
  10.1038/24567} {\bibfield  {journal} {\bibinfo  {journal} {Nature}\ }\textbf
  {\bibinfo {volume} {396}},\ \bibinfo {pages} {345} (\bibinfo {year}
  {1998})}\BibitemShut {NoStop}%
\bibitem [{\citenamefont {Barrett}\ \emph {et~al.}(2001)\citenamefont
  {Barrett}, \citenamefont {Sauer},\ and\ \citenamefont
  {Chapman}}]{Barrett2001}%
  \BibitemOpen
  \bibfield  {author} {\bibinfo {author} {\bibfnamefont {M.~D.}\ \bibnamefont
  {Barrett}}, \bibinfo {author} {\bibfnamefont {J.~A.}\ \bibnamefont {Sauer}},
  \ and\ \bibinfo {author} {\bibfnamefont {M.~S.}\ \bibnamefont {Chapman}},\
  }\href {\doibase 10.1103/PhysRevLett.87.010404} {\bibfield  {journal}
  {\bibinfo  {journal} {Phys. Rev. Lett.}\ }\textbf {\bibinfo {volume} {87}},\
  \bibinfo {pages} {010404} (\bibinfo {year} {2001})}\BibitemShut {NoStop}%
\bibitem [{\citenamefont {Chang}\ \emph {et~al.}(2004)\citenamefont {Chang},
  \citenamefont {Hamley}, \citenamefont {Barrett}, \citenamefont {Sauer},
  \citenamefont {Fortier}, \citenamefont {Zhang}, \citenamefont {You},\ and\
  \citenamefont {Chapman}}]{Chang2004}%
  \BibitemOpen
  \bibfield  {author} {\bibinfo {author} {\bibfnamefont {M.~S.}\ \bibnamefont
  {Chang}}, \bibinfo {author} {\bibfnamefont {C.~D.}\ \bibnamefont {Hamley}},
  \bibinfo {author} {\bibfnamefont {M.~D.}\ \bibnamefont {Barrett}}, \bibinfo
  {author} {\bibfnamefont {J.~A.}\ \bibnamefont {Sauer}}, \bibinfo {author}
  {\bibfnamefont {K.~M.}\ \bibnamefont {Fortier}}, \bibinfo {author}
  {\bibfnamefont {W.}~\bibnamefont {Zhang}}, \bibinfo {author} {\bibfnamefont
  {L.}~\bibnamefont {You}}, \ and\ \bibinfo {author} {\bibfnamefont {M.~S.}\
  \bibnamefont {Chapman}},\ }\href {\doibase 10.1103/PhysRevLett.92.140403}
  {\bibfield  {journal} {\bibinfo  {journal} {Phys. Rev. Lett.}\ }\textbf
  {\bibinfo {volume} {92}},\ \bibinfo {pages} {140403} (\bibinfo {year}
  {2004})}\BibitemShut {NoStop}%
\bibitem [{\citenamefont {Sadler}\ \emph {et~al.}(2006)\citenamefont {Sadler},
  \citenamefont {Higbie}, \citenamefont {Leslie}, \citenamefont
  {Vengalattore},\ and\ \citenamefont {Stamper-Kurn}}]{Sadler2006}%
  \BibitemOpen
  \bibfield  {author} {\bibinfo {author} {\bibfnamefont {L.~E.}\ \bibnamefont
  {Sadler}}, \bibinfo {author} {\bibfnamefont {J.~M.}\ \bibnamefont {Higbie}},
  \bibinfo {author} {\bibfnamefont {S.~R.}\ \bibnamefont {Leslie}}, \bibinfo
  {author} {\bibfnamefont {M.}~\bibnamefont {Vengalattore}}, \ and\ \bibinfo
  {author} {\bibfnamefont {D.~M.}\ \bibnamefont {Stamper-Kurn}},\ }\href
  {\doibase 10.1038/nature05094} {\bibfield  {journal} {\bibinfo  {journal}
  {Nature}\ }\textbf {\bibinfo {volume} {443}},\ \bibinfo {pages} {312}
  (\bibinfo {year} {2006})}\BibitemShut {NoStop}%
\bibitem [{\citenamefont {Luo}\ \emph {et~al.}(2017)\citenamefont {Luo},
  \citenamefont {Zou}, \citenamefont {Wu}, \citenamefont {Liu}, \citenamefont
  {Han}, \citenamefont {Tey},\ and\ \citenamefont {You}}]{Luo2017}%
  \BibitemOpen
  \bibfield  {author} {\bibinfo {author} {\bibfnamefont {X.-y.}\ \bibnamefont
  {Luo}}, \bibinfo {author} {\bibfnamefont {Y.-q.}\ \bibnamefont {Zou}},
  \bibinfo {author} {\bibfnamefont {L.-n.}\ \bibnamefont {Wu}}, \bibinfo
  {author} {\bibfnamefont {Q.}~\bibnamefont {Liu}}, \bibinfo {author}
  {\bibfnamefont {M.-f.}\ \bibnamefont {Han}}, \bibinfo {author} {\bibfnamefont
  {M.~K.}\ \bibnamefont {Tey}}, \ and\ \bibinfo {author} {\bibfnamefont
  {L.}~\bibnamefont {You}},\ }\href {\doibase 10.1126/science.aag1106}
  {\bibfield  {journal} {\bibinfo  {journal} {Science}\ }\textbf {\bibinfo
  {volume} {355}},\ \bibinfo {pages} {620} (\bibinfo {year}
  {2017})}\BibitemShut {NoStop}%
\bibitem [{\citenamefont {Polkovnikov}\ and\ \citenamefont
  {Gritsev}(2008)}]{Polkovnikov2008a}%
  \BibitemOpen
  \bibfield  {author} {\bibinfo {author} {\bibfnamefont {A.}~\bibnamefont
  {Polkovnikov}}\ and\ \bibinfo {author} {\bibfnamefont {V.}~\bibnamefont
  {Gritsev}},\ }\href {\doibase 10.1038/nphys963} {\bibfield  {journal}
  {\bibinfo  {journal} {Nature Physics}\ }\textbf {\bibinfo {volume} {4}},\
  \bibinfo {pages} {477} (\bibinfo {year} {2008})}\BibitemShut {NoStop}%
\bibitem [{\citenamefont {De~Grandi}\ and\ \citenamefont
  {Polkovnikov}(2010)}]{DeGrandi2010}%
  \BibitemOpen
  \bibfield  {author} {\bibinfo {author} {\bibfnamefont {C.}~\bibnamefont
  {De~Grandi}}\ and\ \bibinfo {author} {\bibfnamefont {A.}~\bibnamefont
  {Polkovnikov}},\ }\enquote {\bibinfo {title} {Adiabatic perturbation theory:
  From landau--zener problem to quenching through a quantum critical point},}\
  in\ \href {\doibase 10.1007/978-3-642-11470-0_4} {\emph {\bibinfo {booktitle}
  {Quantum Quenching, Annealing and Computation}}},\ \bibinfo {editor} {edited
  by\ \bibinfo {editor} {\bibfnamefont {A.~K.}\ \bibnamefont {Chandra}},
  \bibinfo {editor} {\bibfnamefont {A.}~\bibnamefont {Das}}, \ and\ \bibinfo
  {editor} {\bibfnamefont {B.~K.}\ \bibnamefont {Chakrabarti}}}\ (\bibinfo
  {publisher} {Springer Berlin Heidelberg},\ \bibinfo {address} {Berlin,
  Heidelberg},\ \bibinfo {year} {2010})\ pp.\ \bibinfo {pages}
  {75--114}\BibitemShut {NoStop}%
\bibitem [{\citenamefont {Law}\ \emph {et~al.}(1998)\citenamefont {Law},
  \citenamefont {Pu},\ and\ \citenamefont {Bigelow}}]{Law1998}%
  \BibitemOpen
  \bibfield  {author} {\bibinfo {author} {\bibfnamefont {C.~K.}\ \bibnamefont
  {Law}}, \bibinfo {author} {\bibfnamefont {H.}~\bibnamefont {Pu}}, \ and\
  \bibinfo {author} {\bibfnamefont {N.~P.}\ \bibnamefont {Bigelow}},\ }\href
  {\doibase 10.1103/PhysRevLett.81.5257} {\bibfield  {journal} {\bibinfo
  {journal} {Phys. Rev. Lett.}\ }\textbf {\bibinfo {volume} {81}},\ \bibinfo
  {pages} {5257} (\bibinfo {year} {1998})}\BibitemShut {NoStop}%
\bibitem [{\citenamefont {Pu}\ \emph {et~al.}(1999)\citenamefont {Pu},
  \citenamefont {Law}, \citenamefont {Raghavan}, \citenamefont {Eberly},\ and\
  \citenamefont {Bigelow}}]{Pu1999}%
  \BibitemOpen
  \bibfield  {author} {\bibinfo {author} {\bibfnamefont {H.}~\bibnamefont
  {Pu}}, \bibinfo {author} {\bibfnamefont {C.~K.}\ \bibnamefont {Law}},
  \bibinfo {author} {\bibfnamefont {S.}~\bibnamefont {Raghavan}}, \bibinfo
  {author} {\bibfnamefont {J.~H.}\ \bibnamefont {Eberly}}, \ and\ \bibinfo
  {author} {\bibfnamefont {N.~P.}\ \bibnamefont {Bigelow}},\ }\href {\doibase
  10.1103/PhysRevA.60.1463} {\bibfield  {journal} {\bibinfo  {journal} {Phys.
  Rev. A}\ }\textbf {\bibinfo {volume} {60}},\ \bibinfo {pages} {1463}
  (\bibinfo {year} {1999})}\BibitemShut {NoStop}%
\bibitem [{\citenamefont {Gaunt}\ \emph {et~al.}(2013)\citenamefont {Gaunt},
  \citenamefont {Schmidutz}, \citenamefont {Gotlibovych}, \citenamefont
  {Smith},\ and\ \citenamefont {Hadzibabic}}]{Gaunt2013}%
  \BibitemOpen
  \bibfield  {author} {\bibinfo {author} {\bibfnamefont {A.~L.}\ \bibnamefont
  {Gaunt}}, \bibinfo {author} {\bibfnamefont {T.~F.}\ \bibnamefont
  {Schmidutz}}, \bibinfo {author} {\bibfnamefont {I.}~\bibnamefont
  {Gotlibovych}}, \bibinfo {author} {\bibfnamefont {R.~P.}\ \bibnamefont
  {Smith}}, \ and\ \bibinfo {author} {\bibfnamefont {Z.}~\bibnamefont
  {Hadzibabic}},\ }\href {\doibase 10.1103/PhysRevLett.110.200406} {\bibfield
  {journal} {\bibinfo  {journal} {Phys. Rev. Lett.}\ }\textbf {\bibinfo
  {volume} {110}},\ \bibinfo {pages} {200406} (\bibinfo {year}
  {2013})}\BibitemShut {NoStop}%
\bibitem [{\citenamefont {Chomaz}\ \emph {et~al.}(2015)\citenamefont {Chomaz},
  \citenamefont {Corman}, \citenamefont {Bienaim{\'{e}}}, \citenamefont
  {Desbuquois}, \citenamefont {Weitenberg}, \citenamefont {Nascimb{\`{e}}ne},
  \citenamefont {Beugnon},\ and\ \citenamefont {Dalibard}}]{Chomaz2015}%
  \BibitemOpen
  \bibfield  {author} {\bibinfo {author} {\bibfnamefont {L.}~\bibnamefont
  {Chomaz}}, \bibinfo {author} {\bibfnamefont {L.}~\bibnamefont {Corman}},
  \bibinfo {author} {\bibfnamefont {T.}~\bibnamefont {Bienaim{\'{e}}}},
  \bibinfo {author} {\bibfnamefont {R.}~\bibnamefont {Desbuquois}}, \bibinfo
  {author} {\bibfnamefont {C.}~\bibnamefont {Weitenberg}}, \bibinfo {author}
  {\bibfnamefont {S.}~\bibnamefont {Nascimb{\`{e}}ne}}, \bibinfo {author}
  {\bibfnamefont {J.}~\bibnamefont {Beugnon}}, \ and\ \bibinfo {author}
  {\bibfnamefont {J.}~\bibnamefont {Dalibard}},\ }\href {\doibase
  10.1038/ncomms7162} {\bibfield  {journal} {\bibinfo  {journal} {Nat.
  Commun.}\ }\textbf {\bibinfo {volume} {6}},\ \bibinfo {pages} {6162}
  (\bibinfo {year} {2015})}\BibitemShut {NoStop}%
\bibitem [{\citenamefont {Beugnon}\ and\ \citenamefont
  {Navon}(2017)}]{Beugnon2016}%
  \BibitemOpen
  \bibfield  {author} {\bibinfo {author} {\bibfnamefont {J.}~\bibnamefont
  {Beugnon}}\ and\ \bibinfo {author} {\bibfnamefont {N.}~\bibnamefont
  {Navon}},\ }\href {\doibase 10.1088/1361-6455/50/2/022002} {\bibfield
  {journal} {\bibinfo  {journal} {J. Phys. B At. Mol. Opt. Phys.}\ }\textbf
  {\bibinfo {volume} {50}},\ \bibinfo {pages} {022002} (\bibinfo {year}
  {2017})}\BibitemShut {NoStop}%
\bibitem [{\citenamefont {Mukherjee}\ \emph {et~al.}(2017)\citenamefont
  {Mukherjee}, \citenamefont {Yan}, \citenamefont {Patel}, \citenamefont
  {Hadzibabic}, \citenamefont {Yefsah}, \citenamefont {Struck},\ and\
  \citenamefont {Zwierlein}}]{Mukherjee2017}%
  \BibitemOpen
  \bibfield  {author} {\bibinfo {author} {\bibfnamefont {B.}~\bibnamefont
  {Mukherjee}}, \bibinfo {author} {\bibfnamefont {Z.}~\bibnamefont {Yan}},
  \bibinfo {author} {\bibfnamefont {P.~B.}\ \bibnamefont {Patel}}, \bibinfo
  {author} {\bibfnamefont {Z.}~\bibnamefont {Hadzibabic}}, \bibinfo {author}
  {\bibfnamefont {T.}~\bibnamefont {Yefsah}}, \bibinfo {author} {\bibfnamefont
  {J.}~\bibnamefont {Struck}}, \ and\ \bibinfo {author} {\bibfnamefont {M.~W.}\
  \bibnamefont {Zwierlein}},\ }\href {\doibase 10.1103/PhysRevLett.118.123401}
  {\bibfield  {journal} {\bibinfo  {journal} {Phys. Rev. Lett.}\ }\textbf
  {\bibinfo {volume} {118}},\ \bibinfo {pages} {123401} (\bibinfo {year}
  {2017})}\BibitemShut {NoStop}%
\bibitem [{\citenamefont {Hueck}\ \emph {et~al.}(2018)\citenamefont {Hueck},
  \citenamefont {Luick}, \citenamefont {Sobirey}, \citenamefont {Siegl},
  \citenamefont {Lompe},\ and\ \citenamefont {Moritz}}]{Hueck2018}%
  \BibitemOpen
  \bibfield  {author} {\bibinfo {author} {\bibfnamefont {K.}~\bibnamefont
  {Hueck}}, \bibinfo {author} {\bibfnamefont {N.}~\bibnamefont {Luick}},
  \bibinfo {author} {\bibfnamefont {L.}~\bibnamefont {Sobirey}}, \bibinfo
  {author} {\bibfnamefont {J.}~\bibnamefont {Siegl}}, \bibinfo {author}
  {\bibfnamefont {T.}~\bibnamefont {Lompe}}, \ and\ \bibinfo {author}
  {\bibfnamefont {H.}~\bibnamefont {Moritz}},\ }\href {\doibase
  10.1103/PhysRevLett.120.060402} {\bibfield  {journal} {\bibinfo  {journal}
  {Phys. Rev. Lett.}\ }\textbf {\bibinfo {volume} {120}},\ \bibinfo {pages}
  {060402} (\bibinfo {year} {2018})}\BibitemShut {NoStop}%
\bibitem [{\citenamefont {Murata}\ \emph {et~al.}(2007)\citenamefont {Murata},
  \citenamefont {Saito},\ and\ \citenamefont {Ueda}}]{Murata2007}%
  \BibitemOpen
  \bibfield  {author} {\bibinfo {author} {\bibfnamefont {K.}~\bibnamefont
  {Murata}}, \bibinfo {author} {\bibfnamefont {H.}~\bibnamefont {Saito}}, \
  and\ \bibinfo {author} {\bibfnamefont {M.}~\bibnamefont {Ueda}},\ }\href
  {\doibase 10.1103/PhysRevA.75.013607} {\bibfield  {journal} {\bibinfo
  {journal} {Phys. Rev. A}\ }\textbf {\bibinfo {volume} {75}},\ \bibinfo
  {pages} {013607} (\bibinfo {year} {2007})}\BibitemShut {NoStop}%
\bibitem [{\citenamefont {Hoang}\ \emph {et~al.}(2016)\citenamefont {Hoang},
  \citenamefont {Bharath}, \citenamefont {Boguslawski}, \citenamefont {Anquez},
  \citenamefont {Robbins},\ and\ \citenamefont {Chapman}}]{Hoang2016a}%
  \BibitemOpen
  \bibfield  {author} {\bibinfo {author} {\bibfnamefont {T.~M.}\ \bibnamefont
  {Hoang}}, \bibinfo {author} {\bibfnamefont {H.~M.}\ \bibnamefont {Bharath}},
  \bibinfo {author} {\bibfnamefont {M.~J.}\ \bibnamefont {Boguslawski}},
  \bibinfo {author} {\bibfnamefont {M.}~\bibnamefont {Anquez}}, \bibinfo
  {author} {\bibfnamefont {B.~A.}\ \bibnamefont {Robbins}}, \ and\ \bibinfo
  {author} {\bibfnamefont {M.~S.}\ \bibnamefont {Chapman}},\ }\href {\doibase
  10.1073/pnas.1600267113} {\bibfield  {journal} {\bibinfo  {journal} {Proc.
  Natl. Acad. Sci.}\ }\textbf {\bibinfo {volume} {113}},\ \bibinfo {pages}
  {9475} (\bibinfo {year} {2016})}\BibitemShut {NoStop}%
\bibitem [{\citenamefont {Botet}\ \emph {et~al.}(1982)\citenamefont {Botet},
  \citenamefont {Jullien},\ and\ \citenamefont {Pfeuty}}]{Botet1982}%
  \BibitemOpen
  \bibfield  {author} {\bibinfo {author} {\bibfnamefont {R.}~\bibnamefont
  {Botet}}, \bibinfo {author} {\bibfnamefont {R.}~\bibnamefont {Jullien}}, \
  and\ \bibinfo {author} {\bibfnamefont {P.}~\bibnamefont {Pfeuty}},\ }\href
  {\doibase 10.1103/PhysRevLett.49.478} {\bibfield  {journal} {\bibinfo
  {journal} {Phys. Rev. Lett.}\ }\textbf {\bibinfo {volume} {49}},\ \bibinfo
  {pages} {478} (\bibinfo {year} {1982})}\BibitemShut {NoStop}%
\bibitem [{\citenamefont {Botet}\ and\ \citenamefont
  {Jullien}(1983)}]{Botet1983}%
  \BibitemOpen
  \bibfield  {author} {\bibinfo {author} {\bibfnamefont {R.}~\bibnamefont
  {Botet}}\ and\ \bibinfo {author} {\bibfnamefont {R.}~\bibnamefont
  {Jullien}},\ }\href {\doibase 10.1103/PhysRevB.28.3955} {\bibfield  {journal}
  {\bibinfo  {journal} {Phys. Rev. B}\ }\textbf {\bibinfo {volume} {28}},\
  \bibinfo {pages} {3955} (\bibinfo {year} {1983})}\BibitemShut {NoStop}%
\bibitem [{\citenamefont {Zhang}\ and\ \citenamefont {Duan}(2013)}]{Zhang2013}%
  \BibitemOpen
  \bibfield  {author} {\bibinfo {author} {\bibfnamefont {Z.}~\bibnamefont
  {Zhang}}\ and\ \bibinfo {author} {\bibfnamefont {L.~M.}\ \bibnamefont
  {Duan}},\ }\href {\doibase 10.1103/PhysRevLett.111.180401} {\bibfield
  {journal} {\bibinfo  {journal} {Phys. Rev. Lett.}\ }\textbf {\bibinfo
  {volume} {111}},\ \bibinfo {pages} {180401} (\bibinfo {year}
  {2013})}\BibitemShut {NoStop}%
\bibitem [{\citenamefont {Vidal}\ and\ \citenamefont
  {Dusuel}(2006)}]{jvidalDickeModel}%
  \BibitemOpen
  \bibfield  {author} {\bibinfo {author} {\bibfnamefont {J.}~\bibnamefont
  {Vidal}}\ and\ \bibinfo {author} {\bibfnamefont {S.}~\bibnamefont {Dusuel}},\
  }\href {http://stacks.iop.org/0295-5075/74/i=5/a=817} {\bibfield  {journal}
  {\bibinfo  {journal} {EPL (Europhysics Letters)}\ }\textbf {\bibinfo {volume}
  {74}},\ \bibinfo {pages} {817} (\bibinfo {year} {2006})}\BibitemShut
  {NoStop}%
\bibitem [{\citenamefont {Dusuel}\ and\ \citenamefont
  {Vidal}(2004)}]{Dusuel2004}%
  \BibitemOpen
  \bibfield  {author} {\bibinfo {author} {\bibfnamefont {S.}~\bibnamefont
  {Dusuel}}\ and\ \bibinfo {author} {\bibfnamefont {J.}~\bibnamefont {Vidal}},\
  }\href {\doibase 10.1103/PhysRevLett.93.237204} {\bibfield  {journal}
  {\bibinfo  {journal} {Phys. Rev. Lett.}\ }\textbf {\bibinfo {volume} {93}},\
  \bibinfo {pages} {237204} (\bibinfo {year} {2004})}\BibitemShut {NoStop}%
\bibitem [{\citenamefont {Leyvraz}\ and\ \citenamefont
  {Heiss}(2005)}]{Leyvraz2005}%
  \BibitemOpen
  \bibfield  {author} {\bibinfo {author} {\bibfnamefont {F.}~\bibnamefont
  {Leyvraz}}\ and\ \bibinfo {author} {\bibfnamefont {W.~D.}\ \bibnamefont
  {Heiss}},\ }\href {\doibase 10.1103/PhysRevLett.95.050402} {\bibfield
  {journal} {\bibinfo  {journal} {Phys. Rev. Lett.}\ }\textbf {\bibinfo
  {volume} {95}},\ \bibinfo {pages} {050402} (\bibinfo {year}
  {2005})}\BibitemShut {NoStop}%
\bibitem [{\citenamefont {Kolodrubetz}\ \emph {et~al.}(2012)\citenamefont
  {Kolodrubetz}, \citenamefont {Pekker}, \citenamefont {Clark},\ and\
  \citenamefont {Sengupta}}]{Kolodrubetz2012b}%
  \BibitemOpen
  \bibfield  {author} {\bibinfo {author} {\bibfnamefont {M.}~\bibnamefont
  {Kolodrubetz}}, \bibinfo {author} {\bibfnamefont {D.}~\bibnamefont {Pekker}},
  \bibinfo {author} {\bibfnamefont {B.~K.}\ \bibnamefont {Clark}}, \ and\
  \bibinfo {author} {\bibfnamefont {K.}~\bibnamefont {Sengupta}},\ }\href
  {\doibase 10.1103/PhysRevB.85.100505} {\bibfield  {journal} {\bibinfo
  {journal} {Phys. Rev. B}\ }\textbf {\bibinfo {volume} {85}},\ \bibinfo
  {pages} {100505} (\bibinfo {year} {2012})}\BibitemShut {NoStop}%
\bibitem [{\citenamefont {Kolodrubetz}\ \emph {et~al.}(2015)\citenamefont
  {Kolodrubetz}, \citenamefont {Katz},\ and\ \citenamefont
  {Polkovnikov}}]{Kolodrubetz2015}%
  \BibitemOpen
  \bibfield  {author} {\bibinfo {author} {\bibfnamefont {M.}~\bibnamefont
  {Kolodrubetz}}, \bibinfo {author} {\bibfnamefont {E.}~\bibnamefont {Katz}}, \
  and\ \bibinfo {author} {\bibfnamefont {A.}~\bibnamefont {Polkovnikov}},\
  }\href {\doibase 10.1103/PhysRevB.91.054306} {\bibfield  {journal} {\bibinfo
  {journal} {Phys. Rev. B}\ }\textbf {\bibinfo {volume} {91}},\ \bibinfo
  {pages} {054306} (\bibinfo {year} {2015})}\BibitemShut {NoStop}%
\bibitem [{\citenamefont {Dutta}\ \emph {et~al.}(2015)\citenamefont {Dutta},
  \citenamefont {Aeppli}, \citenamefont {Chakrabarti}, \citenamefont
  {Divakaran}, \citenamefont {Rosenbaum},\ and\ \citenamefont
  {Sen}}]{dutta2015quantum}%
  \BibitemOpen
  \bibfield  {author} {\bibinfo {author} {\bibfnamefont {A.}~\bibnamefont
  {Dutta}}, \bibinfo {author} {\bibfnamefont {G.}~\bibnamefont {Aeppli}},
  \bibinfo {author} {\bibfnamefont {B.~K.}\ \bibnamefont {Chakrabarti}},
  \bibinfo {author} {\bibfnamefont {U.}~\bibnamefont {Divakaran}}, \bibinfo
  {author} {\bibfnamefont {T.~F.}\ \bibnamefont {Rosenbaum}}, \ and\ \bibinfo
  {author} {\bibfnamefont {D.}~\bibnamefont {Sen}},\ }\href@noop {} {\emph
  {\bibinfo {title} {Quantum Phase Transitions in Transverse Field Spin Models:
  From Statistical Physics to Quantum Information}}}\ (\bibinfo  {publisher}
  {Cambridge University Press},\ \bibinfo {year} {2015})\BibitemShut {NoStop}%
\bibitem [{\citenamefont {Liu}\ \emph {et~al.}(2014)\citenamefont {Liu},
  \citenamefont {Polkovnikov},\ and\ \citenamefont {Sandvik}}]{Liu2014}%
  \BibitemOpen
  \bibfield  {author} {\bibinfo {author} {\bibfnamefont {C.-W.}\ \bibnamefont
  {Liu}}, \bibinfo {author} {\bibfnamefont {A.}~\bibnamefont {Polkovnikov}}, \
  and\ \bibinfo {author} {\bibfnamefont {A.~W.}\ \bibnamefont {Sandvik}},\
  }\href {\doibase 10.1103/PhysRevB.89.054307} {\bibfield  {journal} {\bibinfo
  {journal} {Phys. Rev. B}\ }\textbf {\bibinfo {volume} {89}},\ \bibinfo
  {pages} {054307} (\bibinfo {year} {2014})}\BibitemShut {NoStop}%
\bibitem [{\citenamefont {Gong}\ \emph {et~al.}(2010)\citenamefont {Gong},
  \citenamefont {Zhong}, \citenamefont {Huang},\ and\ \citenamefont
  {Fan}}]{Gong2010}%
  \BibitemOpen
  \bibfield  {author} {\bibinfo {author} {\bibfnamefont {S.}~\bibnamefont
  {Gong}}, \bibinfo {author} {\bibfnamefont {F.}~\bibnamefont {Zhong}},
  \bibinfo {author} {\bibfnamefont {X.}~\bibnamefont {Huang}}, \ and\ \bibinfo
  {author} {\bibfnamefont {S.}~\bibnamefont {Fan}},\ }\href {\doibase
  10.1088/1367-2630/12/4/043036} {\bibfield  {journal} {\bibinfo  {journal}
  {New J. Phys.}\ }\textbf {\bibinfo {volume} {12}},\ \bibinfo {pages} {043036}
  (\bibinfo {year} {2010})}\BibitemShut {NoStop}%
\bibitem [{\citenamefont {{De Grandi}}\ \emph {et~al.}(2011)\citenamefont {{De
  Grandi}}, \citenamefont {Polkovnikov},\ and\ \citenamefont
  {Sandvik}}]{DeGrandi2011}%
  \BibitemOpen
  \bibfield  {author} {\bibinfo {author} {\bibfnamefont {C.}~\bibnamefont {{De
  Grandi}}}, \bibinfo {author} {\bibfnamefont {A.}~\bibnamefont {Polkovnikov}},
  \ and\ \bibinfo {author} {\bibfnamefont {A.~W.}\ \bibnamefont {Sandvik}},\
  }\href {\doibase 10.1103/PhysRevB.84.224303} {\bibfield  {journal} {\bibinfo
  {journal} {Phys. Rev. B}\ }\textbf {\bibinfo {volume} {84}},\ \bibinfo
  {pages} {224303} (\bibinfo {year} {2011})}\BibitemShut {NoStop}%
\bibitem [{\citenamefont {Defenu}\ \emph {et~al.}(2018)\citenamefont {Defenu},
  \citenamefont {Enss},\ and\ \citenamefont {Morigi}}]{Defenu2018}%
  \BibitemOpen
  \bibfield  {author} {\bibinfo {author} {\bibfnamefont {N.}~\bibnamefont
  {Defenu}}, \bibinfo {author} {\bibfnamefont {T.}~\bibnamefont {Enss}}, \ and\
  \bibinfo {author} {\bibfnamefont {G.}~\bibnamefont {Morigi}},\ }\href
  {http://arxiv.org/abs/1805.00008} {\bibfield  {journal} {\bibinfo  {journal}
  {arXiv:1805.00008}\ } (\bibinfo {year} {2018})}\BibitemShut {NoStop}%
\end{thebibliography}%

\end{document}